\documentclass[a4paper,11pt]{article}
\pdfoutput=1 

\usepackage{jheppub} 
\usepackage[T1]{fontenc} 
\usepackage[table]{xcolor}
\usepackage[english]{babel}
\usepackage[utf8]{inputenc}
\usepackage{amsmath}
\usepackage{graphicx}
\usepackage[colorinlistoftodos]{todonotes}
\usepackage{amsmath}
\usepackage{indentfirst}
\usepackage{fancyhdr}
\usepackage{color}
\usepackage{amssymb}
\usepackage{array} 
\usepackage{colortbl}
\newcolumntype{M}[1]{>{\centering\arraybackslash}m{#1}}
\newcolumntype{N}{@{}m{0pt}@{}}

\newcommand{\bea}{\begin{eqnarray}}
\newcommand{\eea}{\end{eqnarray}}
\newcommand{\bi}{\begin{itemize}}
\newcommand{\ei}{\end{itemize}}
\newcommand{\ben}{\begin{enumerate}}
\newcommand{\een}{\end{enumerate}}
\newcommand{\be}{\begin{equation}}
\newcommand{\ee}{\end{equation}}
\newcommand{\ba}{\begin{align}}
\newcommand{\ea}{\end{align}}

\newcommand\vo{{\mathcal{V}}}

\newcommand{\mc}{\mathcal}

\newcommand{\beqa}{\begin{eqnarray}}
\newcommand{\eeqa}{\end{eqnarray}}

\title{Geometrical Destabilisation of Ultra-Light Axions in String Inflation}

\author[a,b]{Michele Cicoli,}
\author[a,b]{Veronica Guidetti,}
\author[a,b]{Francisco G. Pedro}

\affiliation[a]{\small Dipartimento di Fisica e Astronomia, Universit\`a di Bologna, \\ via Irnerio 46, 40126 Bologna, Italy}
\affiliation[b]{\small INFN, Sezione di Bologna, viale Berti Pichat 6/2, 40127 Bologna, Italy}

\emailAdd{michele.cicoli@unibo.it}
\emailAdd{veronica.guidetti2@unibo.it}
\emailAdd{francisco.pedro@bo.infn.it}

\abstract{We perform a detailed analytical and numerical analysis of the multi-field evolution of Fibre Inflation and show that, regardless of the microscopic realisation of the model, the mass-squared of one of the two ultra-light axions becomes always negative during inflation. This implies that the corresponding isocurvature perturbations experience a potential geometrical destabilisation which seems to bring the system away from the perturbative regime. Therefore we conclude that a full understanding of the inflationary evolution of Fibre Inflation can be achieved only via a non-perturbative analysis where the backreaction of tachyonic isocurvature perturbations is properly taken into account.}


\begin{document} 
\maketitle

\section{Introduction}
\label{sec:introduction}

A typical feature of 4D string models is the presence, at tree-level, of a plethora of massless fields, called moduli. Typically these fields acquire mass via supersymmetry breaking effects like non-vanishing background fluxes at semi-classical level, string loops or $\alpha'$ corrections at perturbative level and higher-derivative contributions to the low-energy effective action. 

Some of these moduli are however periodic axion-like fields which enjoy a shift symmetry that is exact at perturbative level \cite{Svrcek:2006yi, Conlon:2006tq}. Hence they become massive only via tiny non-perturbative effects which tend to make them naturally very light, i.e. exponentially lighter than the gravitino mass which sets the mass scale of all the other non-axionic moduli \cite{Arvanitaki:2009fg, Cicoli:2012sz}. Being ultra-light, these axions are perfect candidates for dark radiation \cite{Cicoli:2012aq, Higaki:2012ar} and quintessence fields \cite{Kaloper:2008qs, Panda:2010uq, Choi:1999xn, Cicoli:2018kdo}, and even for cold dark matter via the misalignment mechanism \cite{Hui:2016ltb}.

Another cosmological application of axion-like particles is to act as curvaton fields \cite{Chun:2004gx, Dimopoulos:2005bp}. In fact, during inflation, these ultra-light fields are expected to be much lighter than the Hubble scale. Hence they acquire isocurvature fluctuations which can be converted into standard adiabatic perturbations when the axions decay. If instead the axions are so light that they are still stable, one has to make sure that they do not contribute significantly to dark matter otherwise the amplitude of the isocurvature fluctuations would tend to be larger that the one detected in CMB observations \cite{Ade:2015xua}.

However, axionic isocurvature fluctuations are guaranteed to remain in the perturbative regime only when the field space is flat. On the other hand, when the fields live on a curved manifold, an interesting dynamics can develop. The effective mass-squared of the isocurvature fluctuations receives additional contributions from the Christoffel symbols and the Ricci scalar which can cause their exponential growth \cite{Renaux-Petel:2015mga}, especially if the entropic modes are ultra-light \cite{Cicoli:2018ccr}. This effect could induce a \emph{geometrical destabilisation} of the inflationary trajectory since the growth of the isocurvature perturbations quickly brings the system in the non-perturbative regime. 

The low-energy limit of string compactifications is a 4D supergravity theory which is generically characterised by non-canonical kinetic terms. Hence, as pointed out in \cite{Cicoli:2018ccr}, due to the generic presence of ultra-light axions and a curved field space, string inflationary models might be plagued by geometric destabilisation problems. 

In this paper, we shall first analyse a toy-model with two fields, $\phi_1$ and $\phi_2$. We shall consider $\phi_2$ as massless and $\phi_1$ as a quintessence-like field whose potential is simply a negative exponential. We shall show that, despite the presence of a negatively curved field space, this system does not feature any geometrical destabilisation due to a non-zero turning rate of the underlying bending trajectory which induces a positive contribution to the mass-squared of the isocurvature fluctuations.

We shall then focus on a type IIB inflationary model, Fibre Inflation (FI) \cite{Cicoli:2008gp, Burgess:2016owb}, which is characterised by the presence of two ultra-light axions and a curved field manifold. This model is particularly promising since it is based on an effective rescaling shift symmetry \cite{Burgess:2014tja} and it allows for the construction of globally consistent Calabi-Yau models with inflation and chirality \cite{Cicoli:2016xae, Cicoli:2017axo} and the study of reheating \cite{Cicoli:2018cgu}. Depending on which effects generate the inflationary potential (1-loop open string corrections \cite{Berg:2005ja,Berg:2007wt,Cicoli:2007xp} or higher derivative $\alpha'$ effects \cite{Ciupke:2015msa, Grimm:2017okk}), slightly different FI models can arise \cite{Cicoli:2008gp,Broy:2015zba,Cicoli:2016chb}. However all of them feature a qualitatively similar shape of the inflationary potential characterised by a trans-Planckian plateau which resembles Starobinsky inflation \cite{Starobinsky:1980te} and supergravity $\alpha$-attractors \cite{Kallosh:2013maa, Kallosh:2017wku}. The inflaton field range is around $5$ in Planck units with larger values bounded by the size of the K\"ahler cone \cite{Cicoli:2018tcq}. Thus primordial gravity waves are at the edge of detectability since the tensor-to-scalar ratio turns out to be of order $0.005\lesssim r\lesssim 0.01$. 

We shall first analyse FI models in the limit where the two ultra-light axions are exactly massless and show that the quantum fluctuations of one of these entropic modes always experience an exponential growth. We shall then try to avoid this geometrical destabilisation by turning on a non-zero axionic mass via non-perturbative effects. However we shall find that, in order to obtain a positive mass-squared of the isocurvature modes, these non-perturbative effects have to be of the same order of magnitude of the loop and higher derivative corrections which generate the inflationary potential. The inflationary model therefore changes completely since it becomes intrinsically multi-field. Hence its dynamics should be re-analysed and the predictions for the cosmological observables should be re-derived. We therefore conclude that, if one requires a typical FI dynamics at leading order, there is no way to avoid a tachyonic instability for one of the two ultra-light axions which quickly brings the system away from the perturbative regime. Thus a full understanding of the inflationary dynamics of FI models can be achieved only by relying on non-perturbative techniques which can include the backreaction of dangerous isocurvature modes. We leave this crucial analysis for future work.

This paper is organised as follows. In Sec. \ref{sec:GeomDest} we briefly review the potential geometrical destabilisation of inflation, stressing that heavy fields are stable while the perturbations associated with ultra-light fields can become unstable in negatively curved field manifolds. We then focus instead on specific examples with ultra-light fields, devoting Sec. \ref{SpecEx} to the study of a toy-model with a quintessence-like potential and Sec. \ref{FIdestab} to the more interesting case of type IIB FI models.

\section{Geometrical destabilisation}
\label{sec:GeomDest}

The phenomenon of geometrical destabilisation follows directly from the mass matrix of gauge invariant scalar perturbations:
\be
Q^i\equiv \delta \phi^i+\frac{\dot{\phi}^i}{H}\, \psi\,, 
\ee
where $\phi^i(t,x)=\phi^i(t)+\delta \phi^i (t,x)$ and $\psi(t,x)$ denotes the scalar perturbation to the metric tensor. Let us therefore briefly review how the mass matrix arises in the context of multi-field models of inflation.

In the 2-field models we will be dealing with, it is convenient to project the gauge invariant perturbations $Q^i$ onto normal $Q^N=N_i Q^i$ and parallel $Q^T=T_i Q^i$ components with respect to the background trajectory. $N^i$ are vectors perpendicular to the inflationary trajectory, $T^i$, having unitary 2-norm with respect to the field space metric, $\gamma_{ij}$:
\be
N^i N^j \gamma_{ij}=1\,, \qquad T^i N^j \gamma_{ij}=0\,, \qquad T^i T^j \gamma_{ij}=1\,.
\ee
From the second order action for the perturbations one finds the following equation of motion \cite{Sasaki:1995aw}:\footnote{In our notation $i, j,...$ denote field space directions while capital indices refer to the $T$ and $N$ orthonormal basis.}
\be
\frac{D^2 Q^i}{\partial t^2}+3 H \frac{D Q^i}{\partial t}+\frac{k^2}{a^2}\,Q^i+{M^i}_j Q^j=0\,.
\label{eq:eomPerts}
\ee
The covariant derivatives are defined as:
\be
\frac{D Q^i}{\partial t}=\frac{\partial Q^i}{\partial t}+\Gamma^i_{jk}\dot{\phi}^j Q^k\,,
\ee
and the connections follow from the field space metric $\gamma_{ij}$. The mass matrix in the field basis reads:
\be
{M^i}_j={V^i}_{;j}-R^i_{klj }\dot{\phi}^k \dot{\phi}^l-\frac{1}{a^3}\frac{D}{\partial t}\left( \frac{a^3}{H}\dot{\phi}^i \dot{\phi}_i\right).
\ee
It is convenient to study the perturbations in the $\{ T,N \}$ basis, where one finds an equation of motion of similar form to \eqref{eq:eomPerts} with the covariant derivatives defined in terms of the spin connection (see e.g. \cite{GrootNibbelink:2001qt,Achucarro:2010da} for more details). Focusing on the equation of motion for a single orthogonal perturbation, $Q^N$, one finds that the mass term takes the form:
\be
m_{\perp,\,{\rm eff}}^2 = V_{;\;NN}+ \epsilon  \,  R \,H^2+3\eta_{\perp}^2\,H^2\,,
\label{eq:m2eff}
\ee
where the projection of the covariant derivative is given by:
\be
V_{;\;NN}=\left(V_{,ij}-\Gamma_{ij}^k V_{,k}\right) N^i N^j\,.
\label{eq:CovDer}
\ee 
In \eqref{eq:m2eff} the first two terms depend both on the geometry of the field space and on the scalar potential, while $\eta_\perp$ is related to the inverse of the radius of curvature of the inflationary trajectory in field space and parametrises its non-geodesicity:
\be
\eta_{\perp}=\frac{V_i N^i}{\dot{\phi}_0 H}\qquad\text{with}\qquad \dot{\phi}_0=\sqrt{\gamma_{ij}\dot{\phi}^i\dot{\phi}^j}\,.
\ee
The second term in \eqref{eq:m2eff} depends on the curvature of the 2-dimensional field space $R$, and is the focus of this work. If negative and sufficiently large it can trigger an instability for the isocurvature perturbations by turning their mass-squared negative \cite{Gong:2011uw, Renaux-Petel:2015mga}, in particular for the case of ultra-light fields where the scalar potential contribution to the mass-squared is negligible \cite{Cicoli:2018ccr}. Before delving into the stability analysis of specific models, let us see under which conditions the geometrical instability may arise.

\subsection{Stability of heavy fields}

The destabilisation originally considered in \cite{Renaux-Petel:2015mga} concerned heavy spectator fields during inflation. These are degrees of freedom with a super-Hubble mass, that naively would not play a r\^ole in the low energy dynamics. It was argued that for an arbitrary inflationary potential $V(\phi)$, the heavy field $\chi$ could induce an instability in models of the form:
\be
\frac{\mc{L}}{\sqrt{-g}} = \frac12 f^2\left(\frac{\chi}{\Lambda}\right) \partial_\mu \phi\partial^\mu\phi
+\frac12 \partial_\mu \chi \partial^\mu \chi- V(\phi)-\frac12 \,M^2 \chi^2\,,
\ee
with $M\gg H$ and $f_{\chi\chi}/f >0$ for a certain range of $\Lambda$, the mass parameter setting the scale of the field space curvature. While the zero mode of the heavy field sits at its minimum $\dot{\chi}=\chi=0$, causing $\eta_\perp=0$, the mass-squared of the isocurvature perturbations:
\be
m_{\perp,\,{\rm eff}}^2 = V_{\chi\chi} -2  \frac{f_{\chi\chi}}{f}\,\epsilon H^2\,,
\label{eq:mTach}
\ee
becomes negative due to the curvature term dominating $m_{\perp,\,{\rm eff}}^2$ and causing the corresponding perturbations to grow uncontrollably. This puzzling observation prompted further analysis \cite{Cicoli:2018ccr}, that pointed out that the behaviour of the system relies on the fact that $V(\chi)$ and $f(\chi/\Lambda)$ have common extrema. Furthermore \cite{Cicoli:2018ccr} showed that the background trajectory leading to \eqref{eq:mTach} is classically unstable, thereby providing the correct interpretation for the negative mass-squared found in \cite{Renaux-Petel:2015mga}. Notice that the stability of heavy fields is in agreement with results previously found in models with non-minimal coupling \cite{Kaiser:2012ak, Schutz:2013fua}. Moreover, ref. \cite{Grocholski:2019mot} has recently shown that, even if the initial conditions are tuned such that $\dot{\chi}=\chi=0$, the backreaction of the isocurvature fluctuations shuts off the instability before reaching the non-linear regime. 

Besides showing that there is no instability for kinetically coupled heavy fields, \cite{Cicoli:2018ccr} also put forth the possibility that a negative field space curvature could trigger an instability for massless spectator fields, a situation that is fairly frequent in string models of inflation and that we will discuss in detail below.

\subsection{Potential destabilisation of ultra-light fields}
\label{PotDest}

The most interesting part of our work will be the study of 2-field systems where $\phi_1$ is the inflaton and $\phi_2$ an ultra-light field. In particular we will focus on cases where the ultra-light field is an axion. Assuming that the inflaton can be canonically normalised, the field space metric can be written as:
\be
\gamma_{ij}=\begin{pmatrix}
1 & 0 \\
0 & f^2(\phi_1) 
\end{pmatrix}.
\label{gammaij}
\ee
This class of metrics occurs often in the closed string moduli sector, where the function $f$ might depend explicitly on the inflaton $\phi_1$ while the dependence on the other heavy moduli $\phi_{\rm h}$ is given in terms of their vacuum expectation values: $f=f(\phi_1,\langle\phi_{\rm h}\rangle)$. The 2-field system is described by:
\be
\left\{
\begin{array}{ll}
\ddot{\phi}_1+3H\dot{\phi}_1-f\,f_1 \dot{\phi}_2^2+V_1=0\\[5pt]
\ddot{\phi}_2+3H\dot{\phi}_2+2\frac{f_1}{f}\dot{\phi}_2\dot{\phi}_1+\frac{V_2}{f^2}=0
\end{array}\right.
\ee
and:
\be
T^a=\frac{1}{\dot{\phi}_0}
\begin{pmatrix}
\dot{\phi}_1\\[5pt] \dot{\phi}_2 
\end{pmatrix}\qquad 
N^a=\frac{1}{\dot{\phi}_0}
\begin{pmatrix}
-f\,\dot{\phi}_2\\[5pt] f^{-1}\,\dot{\phi}_1
\end{pmatrix}\,.
\ee
The turning rate of the trajectory reduces to:
\be
\eta_\perp=\frac{1}{2\epsilon H^3}\left(f^{-1}\dot{\phi}_1V_2-f\dot{\phi}_2 V_1\right),
\ee
where we used $\dot{\phi}_0^2= 2 \epsilon H^2$. This implies:
\bea
m_{\perp,\,{\rm eff}}^2 &=& \frac{1}{\dot{\phi}_0^2}\left[(f\dot{\phi}_2)^2 \left(V_{11}+3\frac{V_1^2}{\dot{\phi}_0^2}\right)-2\dot{\phi}_1(f\dot{\phi}_2)\left(\frac{V_{12}}{f}-\frac{f_1}{f}\frac{V_2}{f}+3\frac{V_1V_2}{f\dot{\phi}_0^2}\right)\right. \nonumber \\
&+& \left. \dot{\phi}_1^2\left(\frac{V_{22}}{f^2}+\frac{f_1}{f}V_1+3\frac{V_2^2}{\dot{\phi}_0^2 f^2}\right)\right]-\dot{\phi}_0^2\frac{f_{11}}{f}\,.
\eea
Defining the fraction of kinetic energy carried by $\phi_1$ and $\phi_2$ as $\alpha_1\equiv \frac{\dot{\phi}_1}{\dot{\phi}_0}$ and $\alpha_2\equiv \frac{f\dot{\phi}_2}{\dot{\phi}_0}$ respectively, the entropic mass-squared can be written as: 
\bea
m_{\perp,\,{\rm eff}}^2&=&\alpha_2^2 \left(V_{11}+3\frac{V_1^2}{\dot{\phi}_0^2}\right)-2\alpha_1\alpha_2\left(\frac{V_{12}}{f}-\frac{f_1}{f}\frac{V_2}{f}+3\frac{V_1V_2}{f\dot{\phi}_0^2}\right) \nonumber \\
&+& \alpha_1^2\left(\frac{V_{22}}{f^2}+\frac{f_1}{f}V_1+3\frac{V_2^2}{\dot{\phi}_0^2 f^2}\right)-\dot{\phi}_0^2\frac{f_{11}}{f}\,.
\label{eq:m2effalpha}
\eea
If $\phi_2$ is ultra-light, i.e. $V_2\simeq 0$, (\ref{eq:m2effalpha}) reduces to: 
\be
m_{\perp,\,{\rm eff}}^2 = \alpha_2^2 \left(V_{11}+3\frac{V_1^2}{\dot{\phi}_0^2}\right)+\alpha_1^2\frac{f_1}{f}V_1-\dot{\phi}_0^2\frac{f_{11}}{f} \,.
\label{eq:meffNCanUL}
\ee
In what follows we shall be interested in models where the curvature is constant and negative:
\be
R=-|R| = -2\,\frac{f_{11}}{f}=\text{constant}\,,
\label{eq:eqf}
\ee
which implies:
\be
f(\phi_1)=A_+ \,e^{\lambda\phi_1}+A_-\,e^{-\lambda\phi_1}\qquad \text{with}\qquad \lambda=\sqrt{\frac{|R|}{2}} \,.
\label{fform}
\ee 
In the two special cases with respectively $A_+=0$ or $A_-=0$, the equations of motion become:
\be
\left\{
\begin{array}{ll}
\ddot{\phi}_1+3H\dot{\phi}_1\mp\lambda A_\pm^2\,e^{\pm 2\lambda\phi_1} \dot{\phi}_2^2 + V_1=0 \\ [5pt]
\ddot{\phi}_2+3H\dot{\phi}_2\pm\lambda \dot{\phi}_2\dot{\phi}_1 =0
\end{array}\right.
\ee 
while the effective mass-squared for the isocurvature perturbation simplifies to:
\be 
m_{\perp,\,{\rm eff}}^2 = -\lambda^2\dot{\phi}_0^2 \pm\lambda\alpha_1^2 V_1+\alpha_2^2\left(\frac{3 V_1^2}{\dot{\phi}_0^2}+V_{11}\right). 
\label{eq:meffULFfExp}
\ee 
In the single-field approximation where $\phi_1$ drives inflation while the background value of $\phi_2$ is essentially frozen, i.e. $\alpha_2\ll\alpha_1\simeq 1$, Eq. \eqref{eq:meffULFfExp} can be approximated as:
\be
m_{\perp,\,{\rm eff}}^2 \simeq \lambda\left(\pm V_1- \lambda\dot{\phi}_0^2 \right).
\label{eq:meffULFfExpSimpl}
\ee
The requirement of having a positive mass-squared for the isocurvature perturbation then reduces to $|V_1|>\lambda\dot{\phi}_0^2$ with $V_1 > 0$ for $A_- = 0$ and $V_1 < 0$ for $A_+ = 0$. Using $\dot{\phi}_0^2= 2\epsilon H^2$, and the single-field slow-roll approximations $H^2\simeq V/3$ and $2\epsilon \simeq (V_1/V)^2$, we can easily see that for $\epsilon \ll 1$ and $\lambda\sim\mc{O}(1)$:
\be
\frac{\lambda \dot\phi_0^2}{|V_1|} \simeq \frac{\lambda}{3}\sqrt{2\epsilon} < 1\,.
\label{Import}
\ee
Hence the positivity of the effective mass-squared of the isocurvature perturbation is determined just by the sign of $V_1$ which is the term associated with the metric connection in Eq. \eqref{eq:CovDer}. Interestingly, the Fibre Inflation models which we will discuss in Sec. \ref{FImodels} feature two ultra-light axions, one with $A_+=0$ and the other with $A_-=0$. Hence one of them has necessarily to be geometrically unstable.  

Since the geometrical destabilisation phenomenon is by definition model dependent, we devote the next two sections to the analysis of specific examples. We first look into a simple quintessence-like potential before turning to the string inspired case of Fibre Inflation.

\section{Stability of quintessence-like potentials}
\label{SpecEx}

\subsection{Equations of motion}
\label{EoM}

Exponential potentials can provide the energy density for driving the observed late time accelerated expansion of the universe. Furthermore their simplicity renders them interesting for our purposes as it allows for exact analytic results. Let us therefore focus on the following toy-model involving a quintessence-like field $\phi_1$ and a massless field $\phi_2$ with non-canonical kinetic terms. The metric has the same form as (\ref{gammaij}) with $f = f_0\,e^{-k_1 \phi_1}$ while the scalar potential reads:
\be
V=V_0 \, e^{-k_2\phi_1} \,.
\ee
From \eqref{eq:meffNCanUL} we see that the effective mass-squared of the isocurvature perturbations is:
\be
m_{\perp,\,{\rm eff}}^2 = k_2\,V \left(\alpha_2^2 k_2 \left(1+\frac{3\,V}{\dot{\phi}_0^2}\right)+\alpha_1^2k_1 \right)-k_1^2\dot{\phi}_0^2 \,.
\label{eq::meffSimplest}
\ee
The equations of motion are:
\be
\left\{
\begin{array}{ll}
\ddot{\phi}_1+3H\dot{\phi}_1+k_1 \left(f \dot{\phi}_2\right)^2-k_2 V =0 \\ [5pt]
\ddot{\phi}_2+\left(3H-2k_1\dot{\phi}_1\right)\dot{\phi}_2=0
\end{array}\right.
\ee
which, after trading cosmic time for the number of efoldings $N=\ln a$, can also be rewritten as (the prime superscript denotes derivatives with respect to $N$): 
\be
\label{eq:EOMSstandardphi1phi2}\left\{
\begin{array}{ll}
\phi_1''+(3-\epsilon)\left(\phi_1'-k_2\right) + k_1 \left(f \phi_2'\right)^2=0 \\ [5pt]
\phi_2''+\left(3-\epsilon-2 k_1 \phi_1'\right)\phi_2'= 0 
\end{array}\right.
\ee
The $\phi_2$ equation can be integrated exactly yielding an explicit expression for the velocity of the ultra-light field:
\be
\phi_2'(N) = C\,e^{-3 N + 2 k_1 \phi_1(N) + \int_0^N \epsilon(\tilde{N}) \,d\tilde{N}}\,,
\ee
where $C = \phi_2'(0)\, e^{- 2 k_1 \phi_1(0)}$. Since the kinetic terms of the massless field $\phi_2$ are non-canonical, it is more appropriate to consider the quantity:
\be
\left(f\phi_2'\right)(N) = f_0 C\,e^{-3N + k_1 \phi_1(N) + \int_0^N \epsilon(\tilde{N})\, d\tilde{N}} \,,
\label{eq:standardfphi2}
\ee
which enters into the inflationary $\epsilon$ parameter:
\be
\epsilon=\frac12\,\phi_1'^2+\frac12 \left(f \phi_2'\right)^2 \,.
\ee
Let us now study the behaviour of the system using both an analytical and a computational approach. In the attractor regime where $\phi_1''\simeq \left(f \phi_2'\right)'\simeq 0$, the equations of motion take the form:
\be
\left\{
\begin{array}{ll}
(3-\epsilon)\left(\phi_1'-k_2\right)+k_1 (f\phi_2')^2=0 \\ [5pt]
\left(3-\epsilon-k_1\phi_1'\right)(f\phi_2')=0 
\end{array}\right.
\ee
The system admits two different solutions depending on whether $\phi_2$ is frozen or not.

\begin{figure}[!h]
\centering
\includegraphics[scale=0.45]{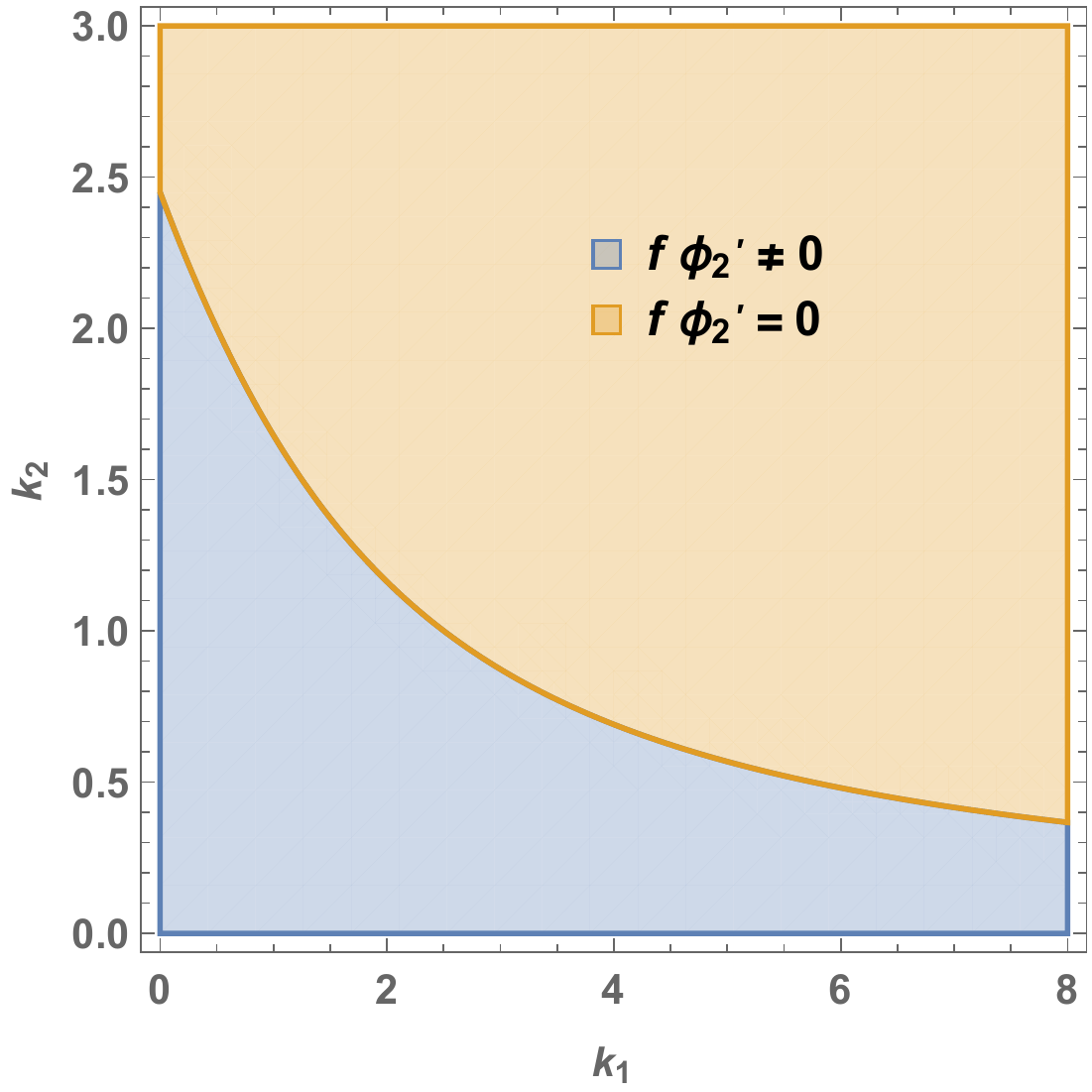}
\caption{Velocity of the massless isocurvature mode in the $(k_1, k_2)$ parameter space.}
\label{Fig1}
\end{figure}

\subsection{Case I: Non-zero turning rate}

This case is characterised by a rolling massless field with $f\phi_2'\ne 0$, $\epsilon=\frac{3 k_2}{(2k_1+k_2)}$ and:
\be
\begin{array}{ll}
\left\{
\begin{array}{lll}
\phi_1'=\frac{3-\epsilon}{k_1} \\ [5pt]
\left(f \phi_2'\right)^2=\phi_1'^2 \left(\frac{k_1k_2}{3-\epsilon}-1\right) 
\end{array}\right.\ .
\end{array}
\ee
For $k_1>0$, the conditions $\epsilon < 1$, $k_1\phi_1'>0$ and $(f \phi_2')^2\geq 0$ can be satisfied only for:
\be
k_2\geq \sqrt{k_1^2+6}-k_1\,.
\ee
It is easy to realise that under this condition the effective mass-squared of the isocurvature mode remains always non-negative:
\be
\frac{m_{\perp,\,{\rm eff}}^2}{H^2}=\frac{6 k_1 (k_2^2 + 2 k_1k_2-6)}{2 k_1+k_2}\geq 0\,.
\ee
The absence of geometrical destabilisation is due to the fact that the trajectory deviates from a simple geodesic since:
\be
\eta_\perp^2 =\left[\frac{(3-\epsilon)}{2\epsilon}\frac{V_1}{V} (f \phi_2')\right]^2 =\frac{k_1}{2k_1+k_2} \frac{m_{\perp,\,{\rm eff}}^2}{H^2}\ne 0\,.
\ee
Notice that in the limiting case where $k_2=-k_1+\sqrt{k_1^2+6}$, the system evolves towards the attractor solution where $f \phi_2'=0$, $\alpha_2=0$, $\alpha_1=1$, $m_{\perp,\,{\rm eff}}^2=0$ and $\eta_\perp=0$. However, we checked that the convergence to this point is extremely slow and the turning rate of the trajectory remains non-negligible for a large number of e-foldings.

\subsection{Case II: Geodesic motion} 

In this case $f\phi_2'=0$, $\epsilon= k_2^2/2$ and the asymptotic state reached by the system is:
\be
\left\{
\begin{array}{lll}
\phi_1'=k_2 \\ [5pt]
(f \phi_2')^2=0
\end{array}\right.
\ee
under the requirement:
\be
k_2 \leq \sqrt{k_1^2+6}-k_1\,.
\ee

\begin{figure}[!h]
  \centering
  \begin{minipage}[b]{0.29\textwidth}
    \includegraphics[width=\textwidth]{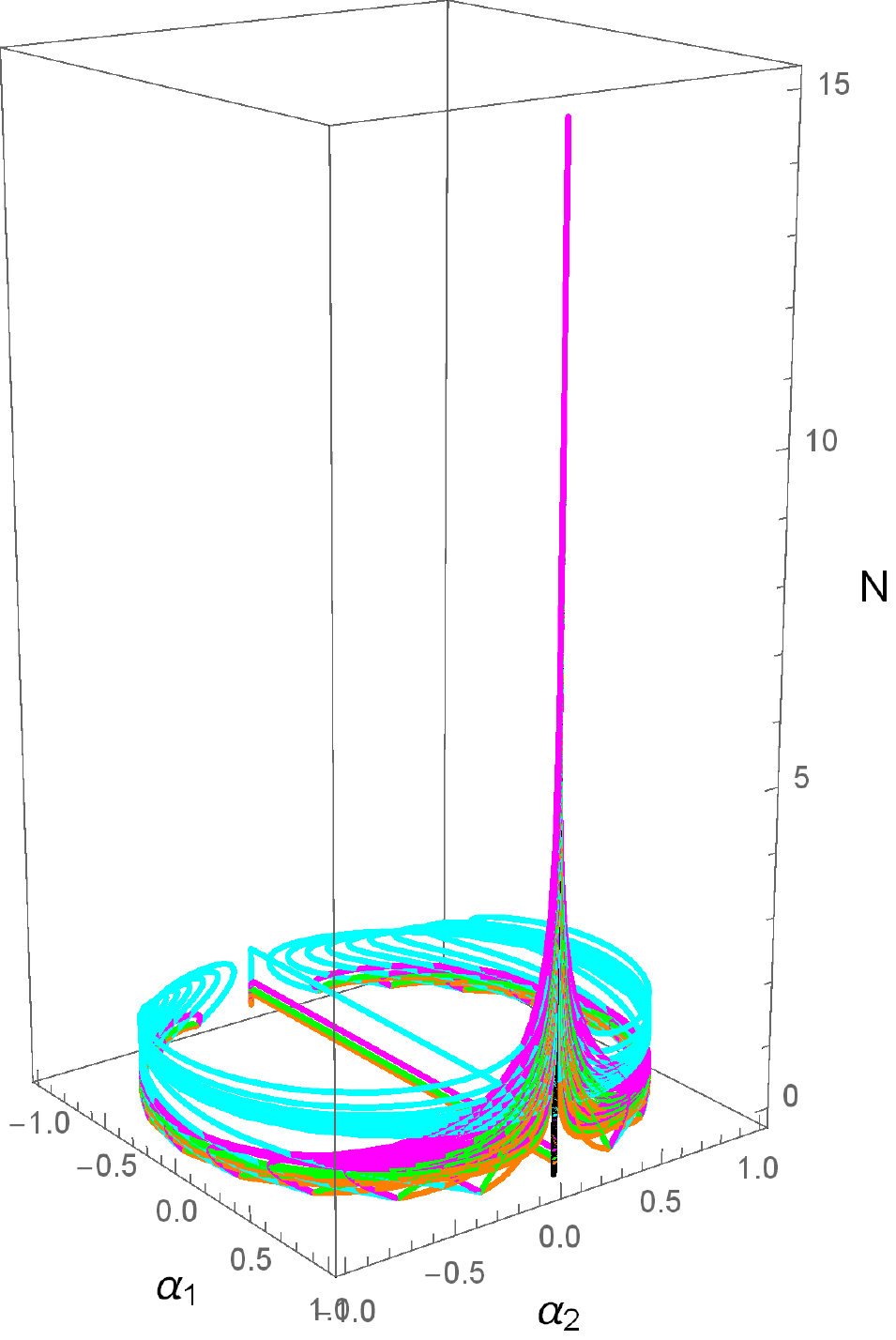}
  \end{minipage}
  \hfill
  \begin{minipage}[b]{0.29\textwidth}
    \includegraphics[width=\textwidth]{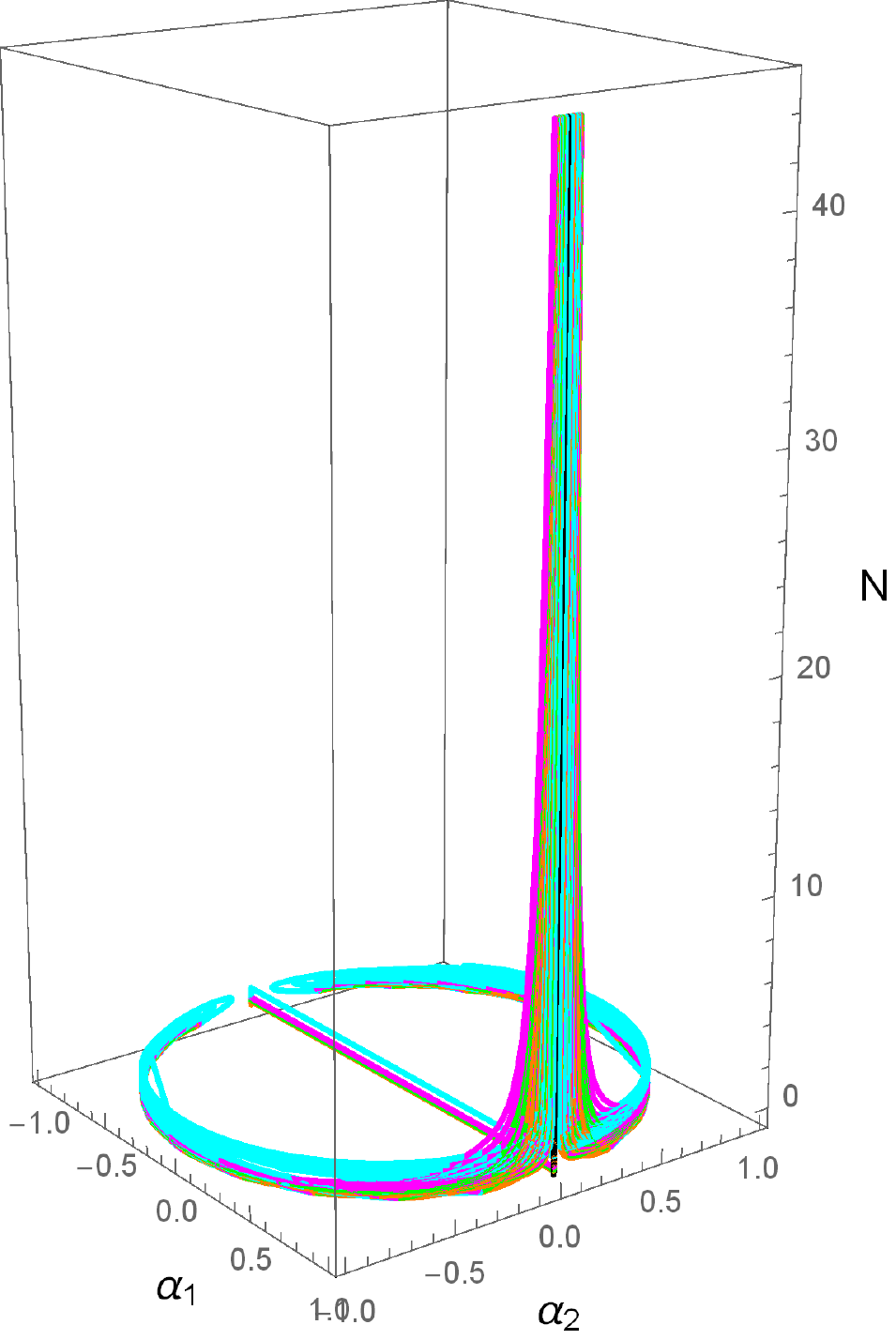}
  \end{minipage}\hfill
  \begin{minipage}[b]{0.37\textwidth}
    \includegraphics[width=\textwidth]{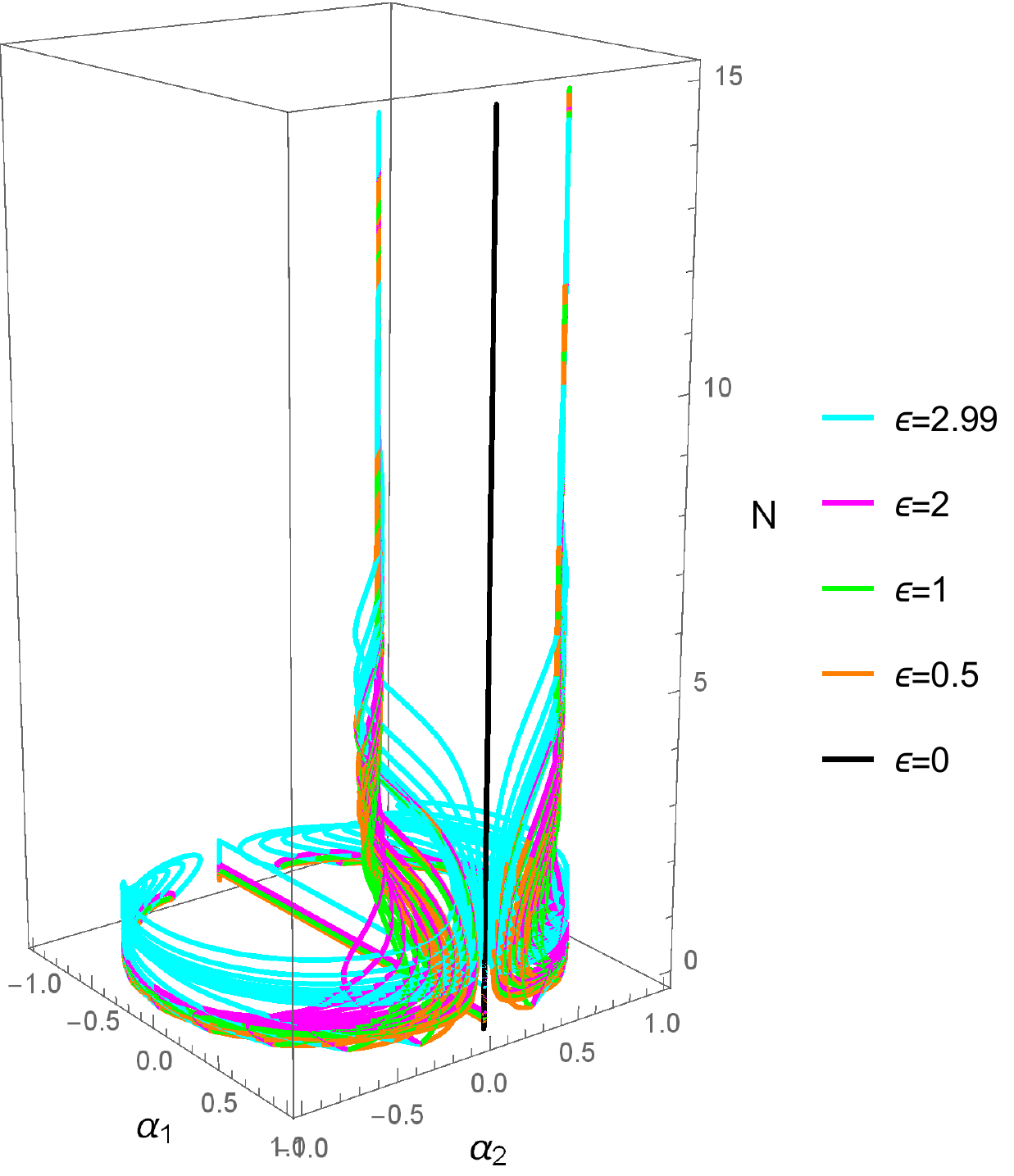}
  \end{minipage}
\caption{Evolution of $\alpha_1\equiv \dot{\phi}_1/\dot{\phi}_0$ and $\alpha_2\equiv f\dot{\phi}_2/\dot{\phi}_0$ (with $\alpha_1^2+\alpha_2^2=1$) for different initial values of $\epsilon=\dot{\phi}_0^2/(2 H^2)$. We set $k_2=\beta(\sqrt{k_1^2+6}-k_1)$ with $\beta=0.8$ (left), $\beta=1$ (centre), $\beta=1.2$ (right) and $k_1=1$.}
\label{fig:alphasbehaviourquint}
\end{figure}

Fig. \ref{Fig1} shows the behaviour of the velocity of the massless field $\phi_2$ for different values of the parameters $k_1$ and $k_2$. In this case the system evolves along a geodesic with $\eta_\perp = 0$ and the sign of the effective mass-squared of the entropic perturbation depends on the sign of $k_2$ since:
\be
\frac{m_{\perp,\,{\rm eff}}^2}{H^2}= \frac{k_1k_2}{2}\underbrace{\left(6-2k_1k_2-k_2^2\right)}_{\geq 0}\,.
\ee

Hence $m_{\perp,\,{\rm eff}}^2\geq 0$ for $k_2>0$, while $m_{\perp,\,{\rm eff}}^2\leq 0$ for $k_2<0$. Notice that in this case geometrical destabilisation can be avoided for $k_2>0$ due to the positive contribution coming from the metric connection. These results are completely independent on the initial conditions.

\subsection{Numerical analysis}

In order to strengthen our analytical results, we also performed a numerical analysis using several parameter sets. We considered different values of the initial kinetic energy $\epsilon_i(0)=\{0,0.5,1,2,3\}$\footnote{These values of $\epsilon$ describe initial conditions ranging from slow-roll ($\epsilon\ll1$) to kinetic domination ($\epsilon=3$). For kinetic domination we actually chose $\epsilon=2.99$ in order to avoid a singularity in the equations of motion stemming from the use of $N$ as the time variable.} and for each of these values we analysed 20 different types of initial conditions for the field velocities:
\be
\left\{
\begin{array}{ll}
\left.\phi_1'(0)\right|_{(ik)} = \sqrt{2 \epsilon_i(0)} \,\cos\left(\frac{k\pi}{10}\right) \\ [5pt]
\left.\left(f\phi_2'\right)(0)\right|_{(ik)} = \sqrt{2 \epsilon_i(0)} \,\sin\left(\frac{k\pi}{10}\right) 
\end{array}\right. \qquad k=0,\dots,19\,.
\label{InCond}
\ee

\begin{figure}[!h]
  \centering
  \begin{minipage}[b]{0.29\textwidth}
    \includegraphics[width=\textwidth]{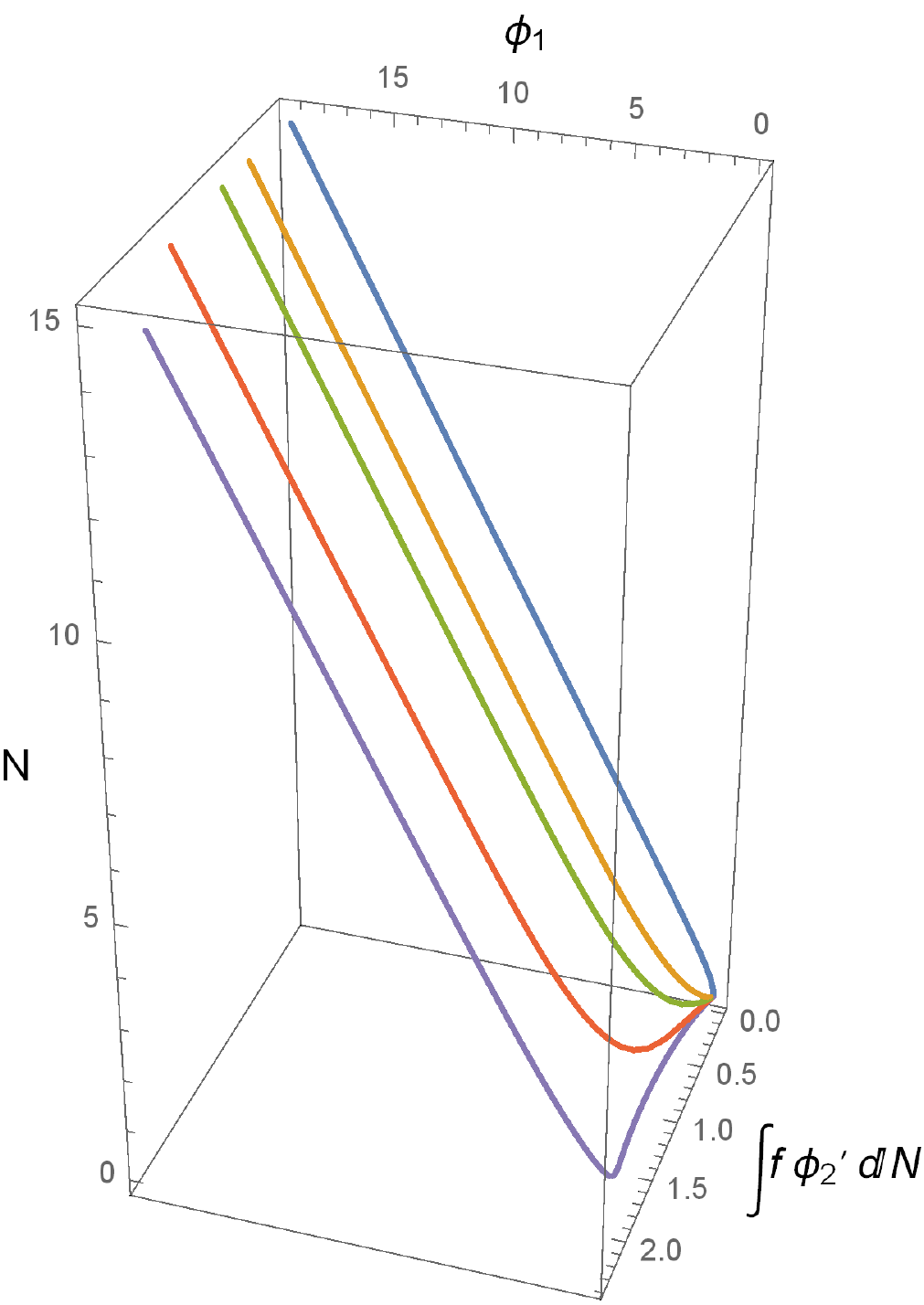}
  \end{minipage}
 \hfill
  \begin{minipage}[b]{0.29\textwidth}
    \includegraphics[width=\textwidth]{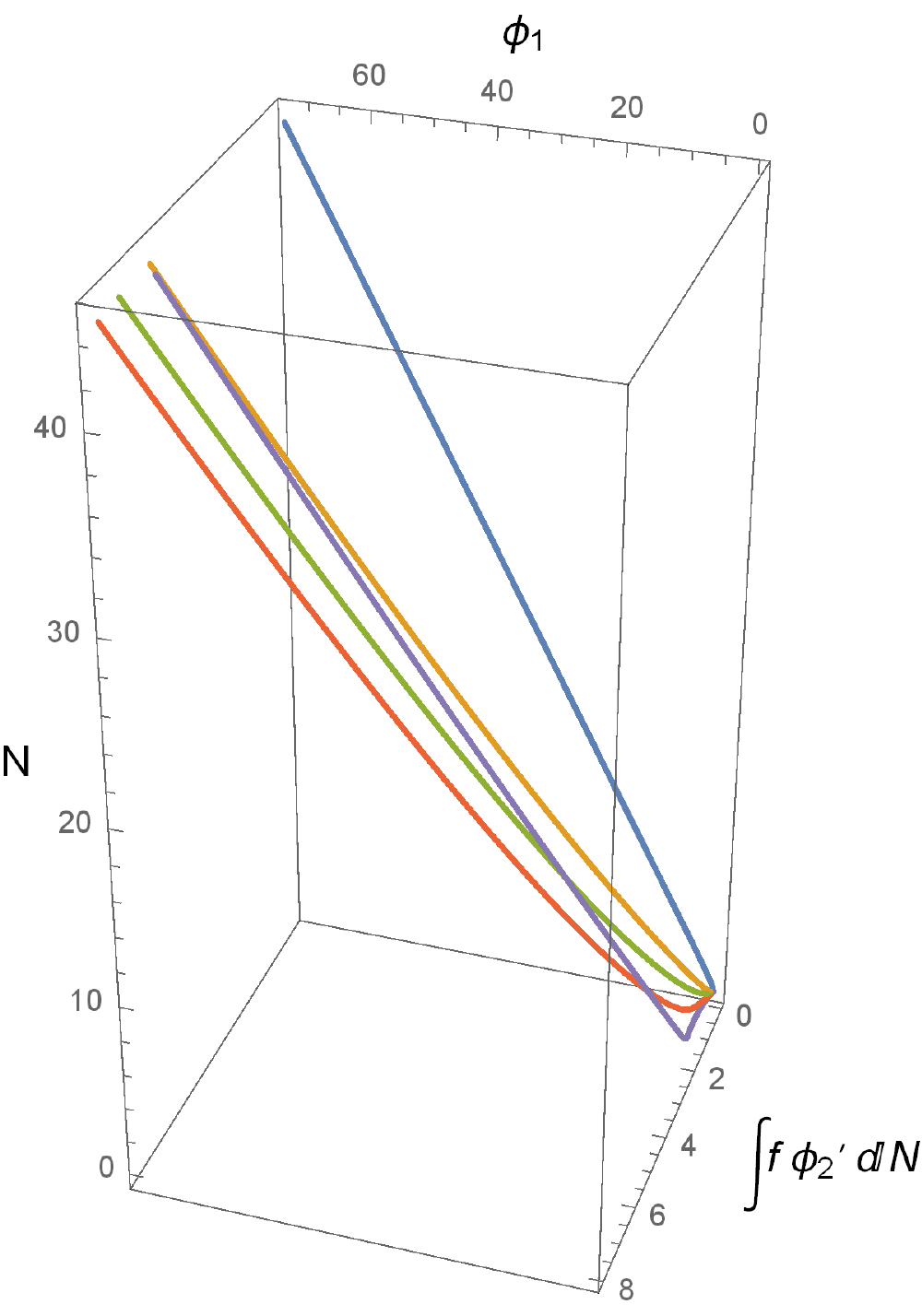}
  \end{minipage}
   \hfill
  \begin{minipage}[b]{0.38\textwidth}
    \includegraphics[width=\textwidth]{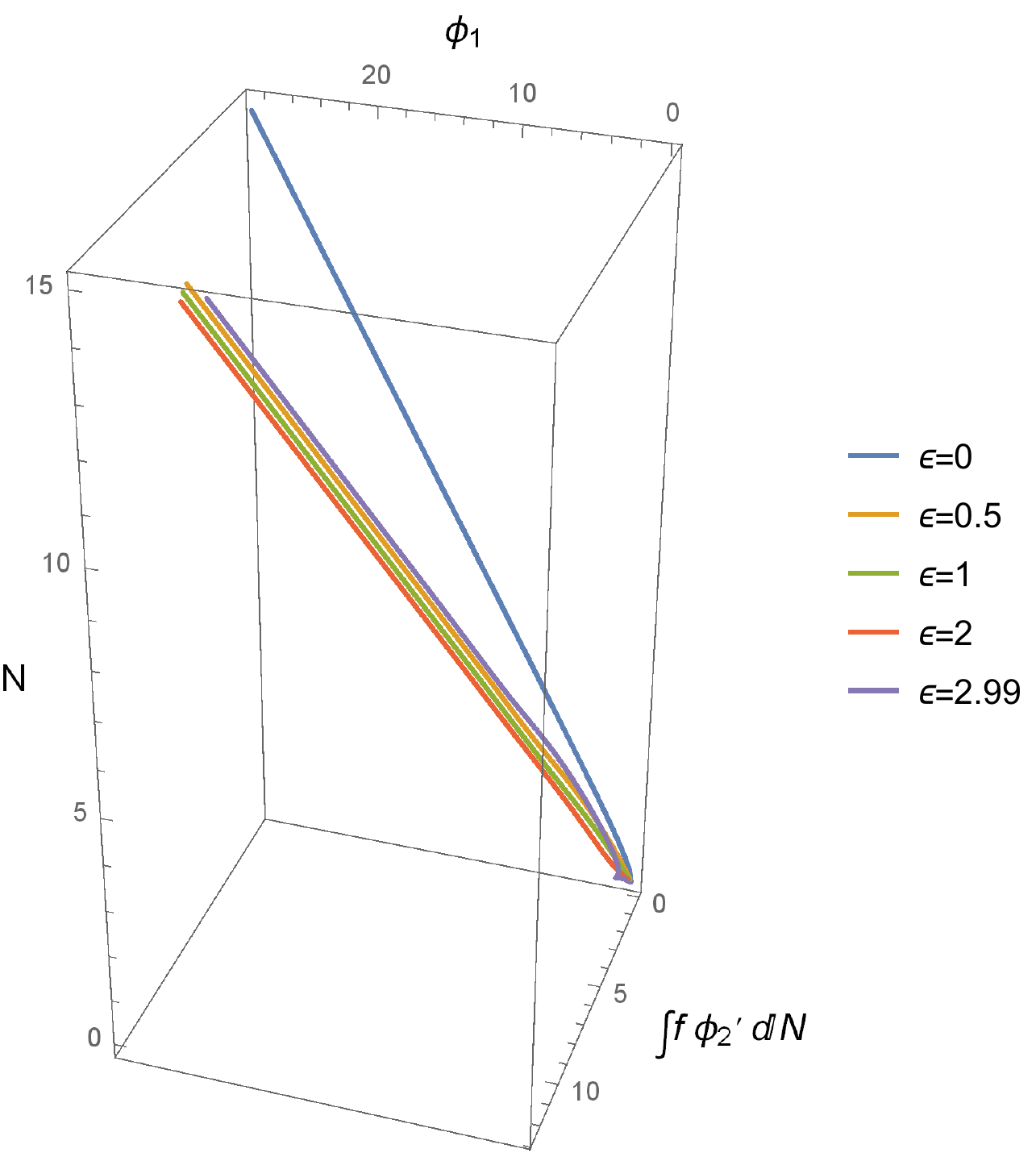}
  \end{minipage}
  \caption{Evolution of the physical fields $\{\phi_1(N),\int_0^N{f(\phi_1(\tilde{N}))\phi_2'(\tilde{N})d\tilde{N}}\}$ for different values of the initial kinetic energy (setting $k=1$ in the initial conditions (\ref{InCond}) for the field velocities) and $\beta=0.8$ (left), $\beta=1$ (centre), $\beta=1.2$ (right).}
  \label{fig:alphasandfields}
\end{figure}

\begin{figure}[!h]
  \centering
  Case I ($\beta=1.2$)\par\bigskip
  \begin{minipage}[b]{0.42\textwidth}
    \includegraphics[width=\textwidth]{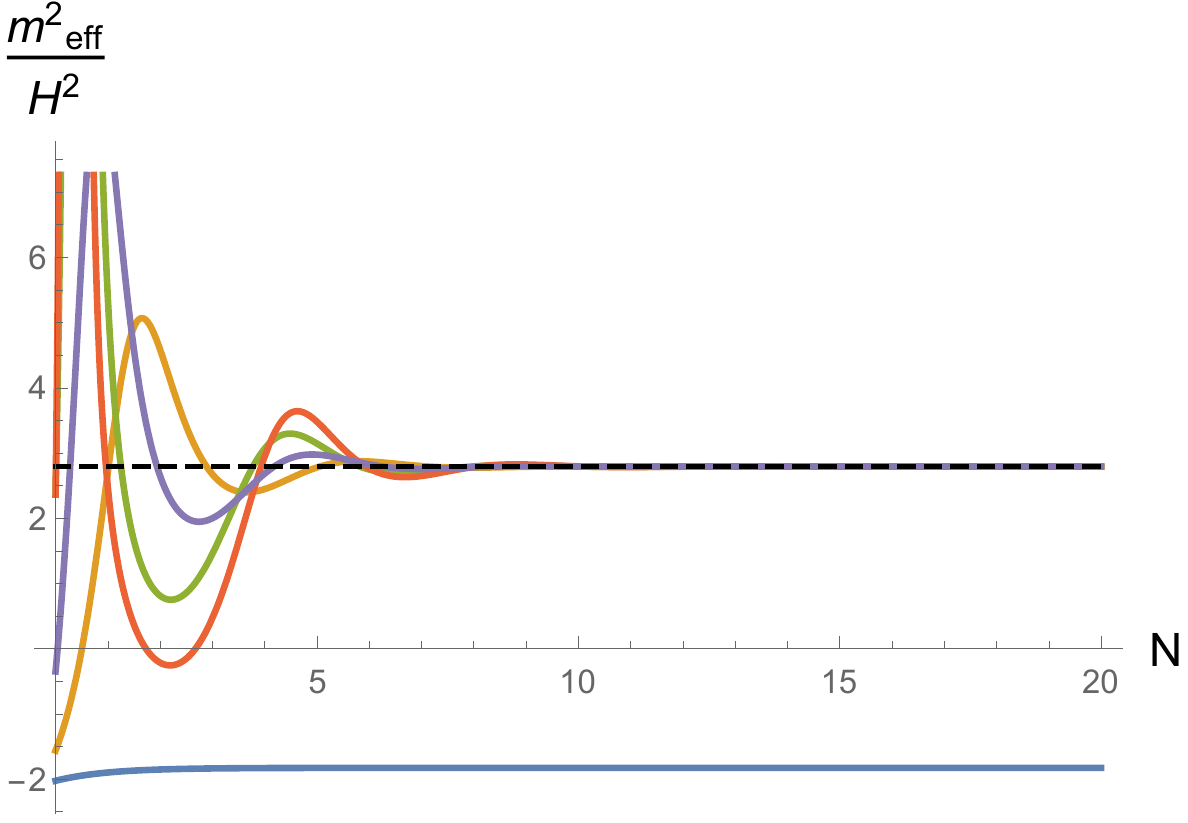}
  \end{minipage}
  \hfill
  \begin{minipage}[b]{0.56\textwidth}
    \includegraphics[width=\textwidth]{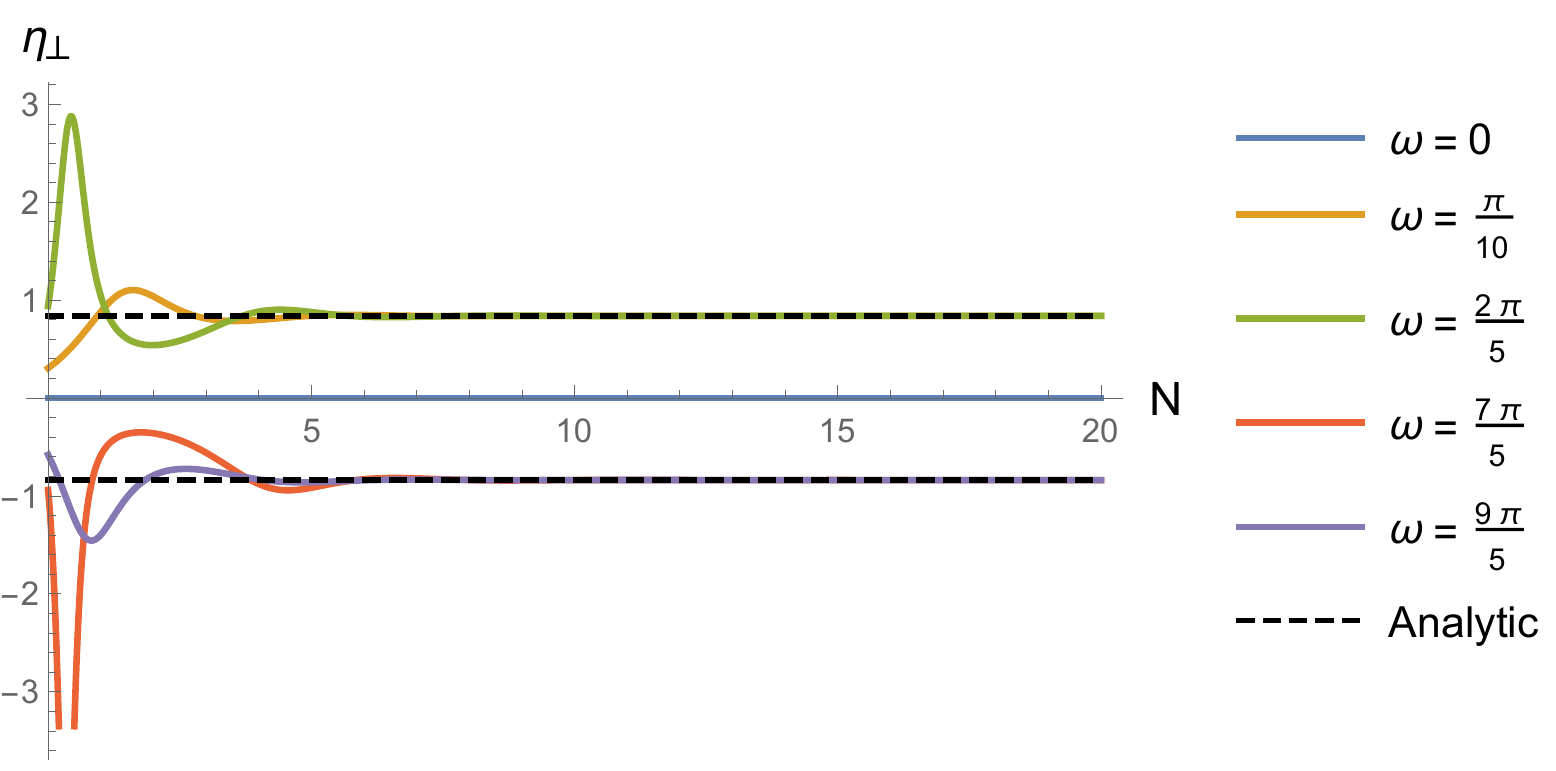}
  \end{minipage}\\[10pt] 
  \centering
  Limiting case ($\beta=1$)\par\bigskip
  \begin{minipage}[b]{0.42\textwidth}
    \includegraphics[width=\textwidth]{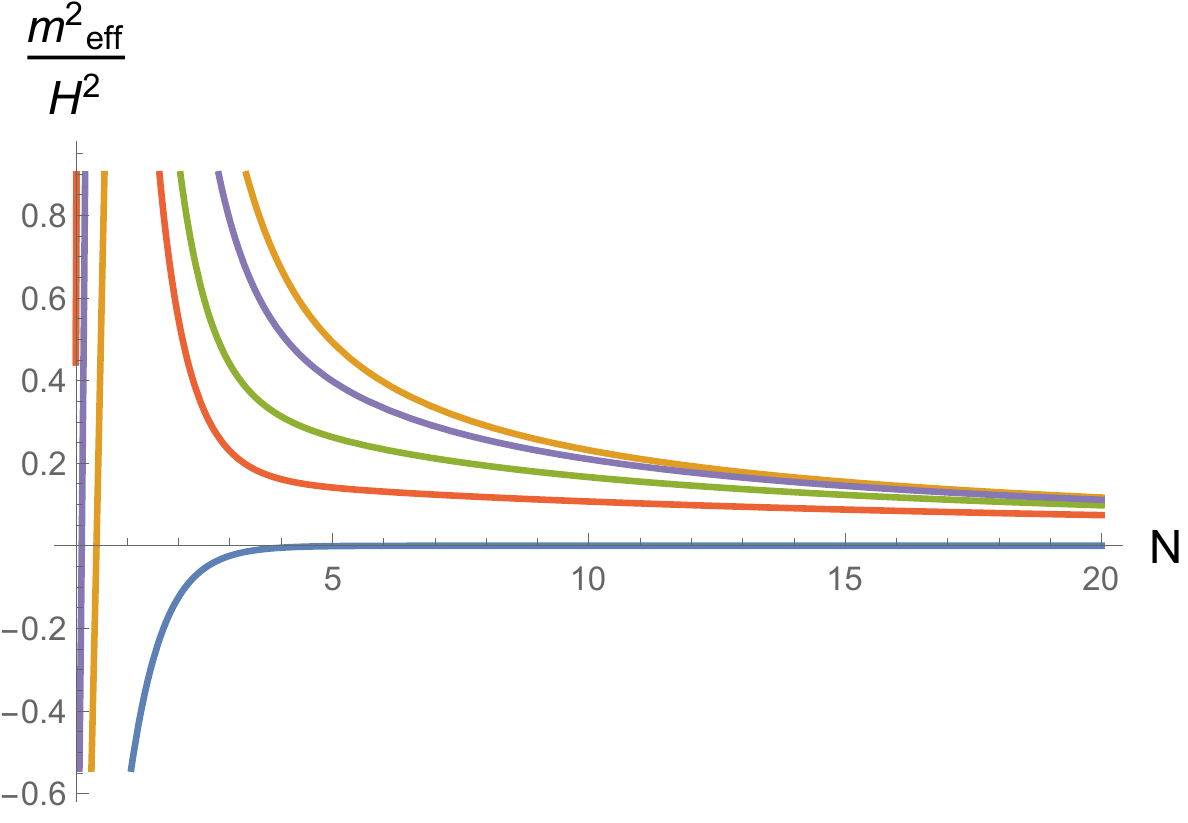}
  \end{minipage}
 \hfill
  \begin{minipage}[b]{0.56\textwidth}
    \includegraphics[width=\textwidth]{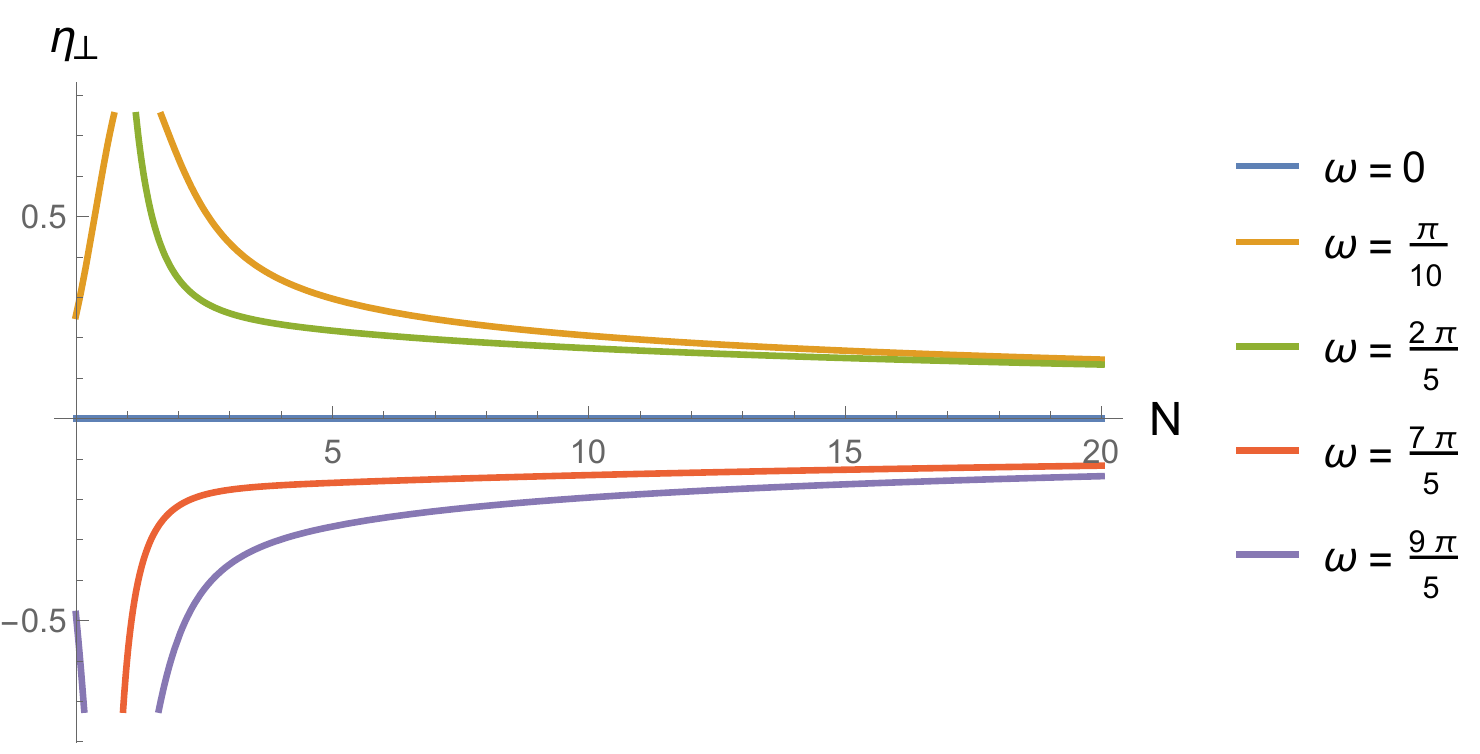}
  \end{minipage}\\[10pt] 
  \centering
  Case II ($\beta=0.8$)\par\bigskip
  \begin{minipage}[b]{0.42\textwidth}
    \includegraphics[width=\textwidth]{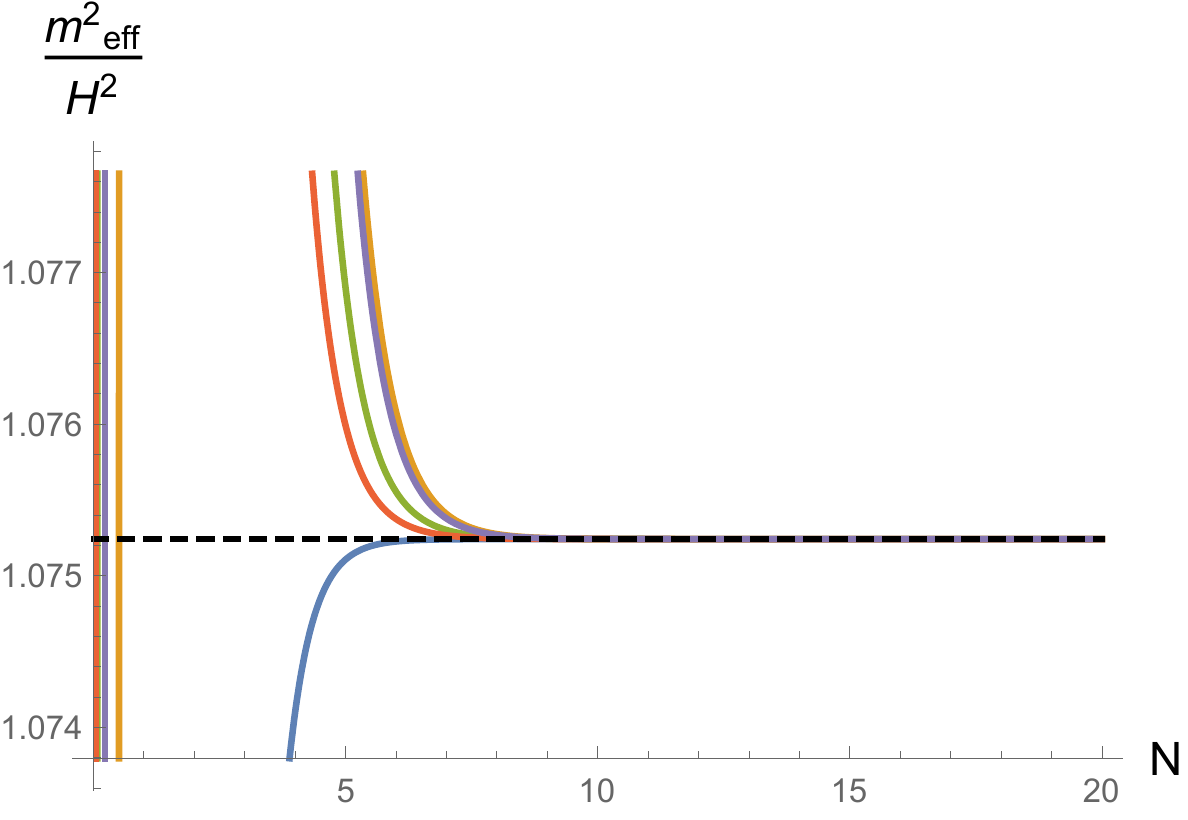}
  \end{minipage}
 \hfill
  \begin{minipage}[b]{0.56\textwidth}
    \includegraphics[width=\textwidth]{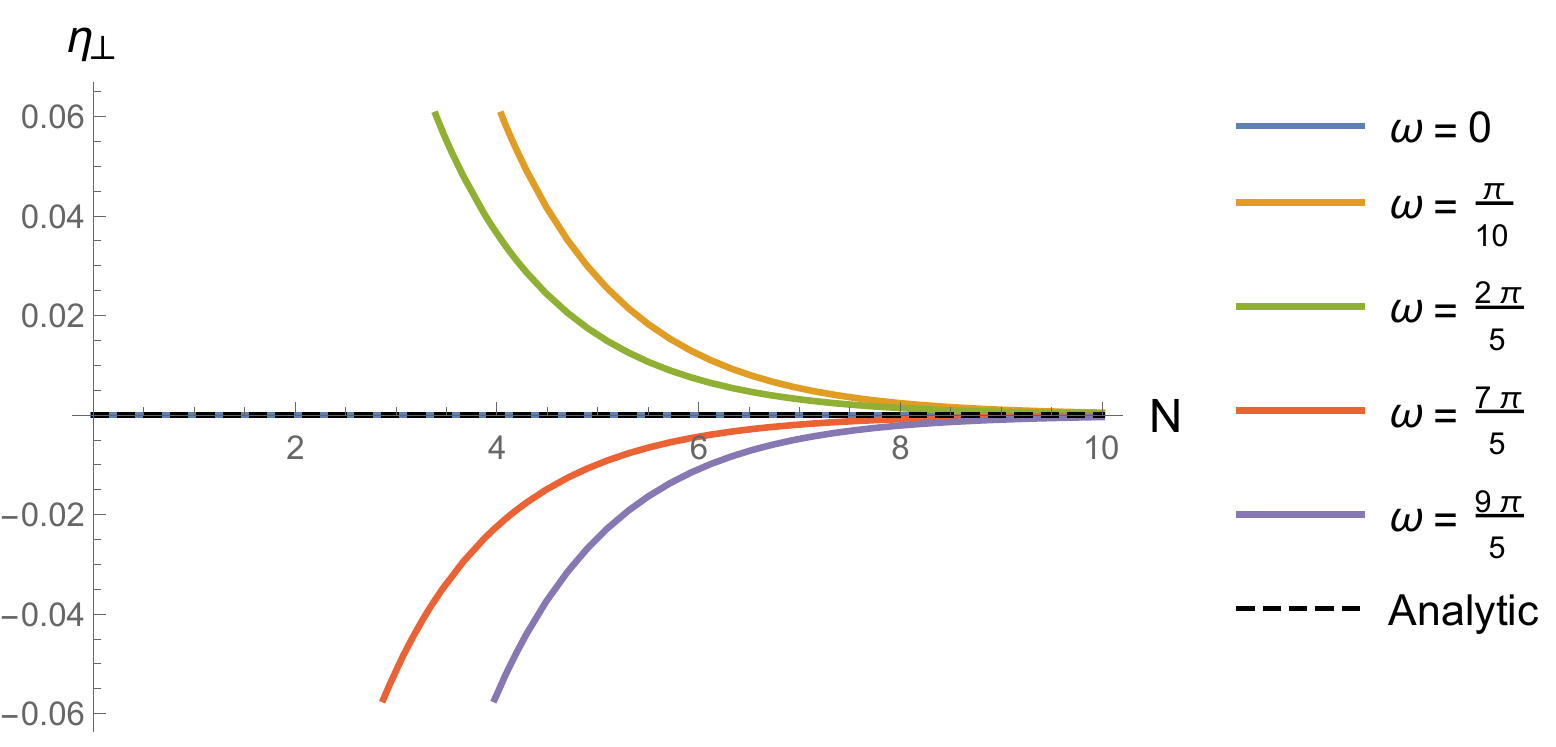}
  \end{minipage}
  \caption{Evolution of $m_{\perp,\,{\rm eff}}^2/H^2$ (left) and $\eta_\perp$ (right) for $\epsilon(0)=2$ and different initial field velocities identified by $\omega_k \equiv k\pi/10$ with $\kappa=\{0,1,4,14,18\}$. The $3$ different cases correspond to $\beta=\{0.8,1,1.2\}$.}
  \label{fig:meffandeta}
\end{figure}

The dynamics of the system is independent on both the initial field values and the normalisation of the scalar potential and the kinetic function. Hence we set, without loss of generality, $\phi_1(0)=\phi_2(0)=0$ and $V_0=f_0=1$. We studied the $3$ interesting cases with $k_2=\beta(\sqrt{k_1^2+6}-k_1)$ and $\beta={0.8,1,1.2}$ for $k_1=1$. Our numerical results are shown in Figs. \ref{fig:alphasbehaviourquint}-\ref{fig:meffandeta} and are in perfect agreement with our analytical analysis. In Fig. \ref{fig:alphasbehaviourquint} we can clearly see that for $\beta\leq 1$ the system converges to a single-field behaviour with $\eta_\perp=0$, while for $\beta>1$ the turning rate of the trajectory is non-zero and the asymptotic behaviour of the system depends on the initial condition for the velocity of the massless field $\phi_2$. 

Notice that the trajectories which move away from the unit circle $\alpha_1^2+\alpha_2^2=1$ correspond to cases with special initial conditions, $\phi_1'(0)$ with the same sign as $V_1$ and $\phi_2'(0)=0$, where $\phi_1$ initially climbs up the potential and then it slows down until it stops and changes its direction. At this point $\phi_1'=\phi_2'=0$, and so the coordinates $\alpha_1$ and $\alpha_2$ are ill-defined. As soon as $\phi_1$ changes its direction, $\phi_1'\neq 0$, and so the system jumps to the opposite point in the unit circle.

Fig. \ref{fig:meffandeta} presents the evolution of $m_{\perp,\,{\rm eff}}^2$ and $\eta_\perp$, showing that the numerical solutions correctly approach our analytic results for $3$ different cases. Notice that in the limiting case with $\beta=1$ all curves tend asymptotically to $m_{\perp,\,{\rm eff}}^2=\eta_\perp=0$, while in case I with $\beta=1.2$, only the blue curve corresponding to $\omega=0$ features a negative mass-squared due to the initial condition $\left(f\phi_2'\right)(0)=0$. This is the only case which could be plagued by a geometric destabilisation problem but it corresponds to a very non-generic choice of initial conditions as argued in \cite{Cicoli:2018ccr}. As soon as $\left(f\phi_2'\right)(0)$ or $\omega$ slightly deviates from zero, Fig. \ref{fig:meffandeta} clearly shows that the mass-squared becomes positive due to a non-vanishing $\eta_\perp$.
  
\section{Geometrical destabilisation in Fibre Inflation}
\label{FIdestab}

\subsection{Fibre Inflation in a nutshell}
\label{FImodels}

The simplest version of Fibre Inflation involves $3$ type IIB K\"ahler moduli $T_i = \tau_i + {\rm i}\theta_i$, $i=1,2,3$ where the $\tau$'s control volumes of $4$-cycles while the $\theta$'s are periodic axion-like fields which enjoy a perturbative shift symmetry. The K\"ahler potential reads:
\be 
K = -2 \ln{\left(\vo+\frac{\xi}{2 g_s^{3/2}}\right)}+ K_{g_s} \,,
\label{KahlerPot}
\ee 
where $\vo = \alpha\left( \sqrt{\tau_1}\tau_2 -\lambda_3 \tau_3^{3/2} \right)$ is the Calabi-Yau volume, the $\mc{O}(1)$ constant $\xi$ controls the leading order $\alpha'$ contribution \cite{Becker:2002nn}, while $K_{g_s}$ denotes 1-loop open string corrections \cite{Berg:2005ja,Berg:2007wt,Cicoli:2007xp}. The superpotential is instead given by a tree-level constant $W_0$ and non-perturbative effects from gaugino condensation on D7-branes or ED3-instantons \cite{Blumenhagen:2009qh}: 
\be
W=W_0+ A_3\, e^{-a_3T_3}\,.
\label{Wsup}
\ee
If $g_s$ loops are neglected, the K\"ahler potential (\ref{KahlerPot}) and the superpotential (\ref{Wsup}) generate a scalar potential of the LVS form \cite{Balasubramanian:2005zx, Cicoli:2008va}:
\be
V_{\scriptscriptstyle \rm LVS} = \frac{8 a_3^2 A_3^2 \sqrt{\tau_3}}{3 \alpha \lambda_3 \vo} e^{-2 a_3 \tau_3}+\frac{4 a_3 A_3 \tau_3 W_0  \cos (a_3 \theta_3)}{\vo^2}\,e^{-a_3 \tau_3} +\frac{3\xi W_0^2}{4 g_s^{3/2}\vo^3}\,,
\label{VLVS}
\ee
which leads to the existence of AdS vacua at exponentially large volume (in string units) where $\vo$, $\tau_3$ and $\theta_3$ are stabilised at:
\be
a_3\langle\theta_3\rangle=\pi\,, \qquad a_3 \langle\tau_3\rangle = \left(\frac{\xi}{2\alpha \lambda_3}\right)^{2/3}\frac{1}{g_s}\,,
\qquad \langle \vo\rangle=\frac{3 W_0 \alpha \lambda_3}{4 a_3 A_3}\,\sqrt{\langle\tau_3\rangle}\,e^{a_3\langle\tau_3\rangle}\,. 
\ee
It is easy to see that the scalar potential (\ref{VLVS}) features three flat directions corresponding to $\tau_1$ and the two axions $\theta_1$ and $\theta_2$. The inclusion of subleading $g_s$ or $\alpha'$ corrections to the K\"ahler potential can lift $\tau_1$ but not $\theta_1$ and $\theta_2$ which are protected by a perturbative shift symmetry. In the presence of sources of positive vacuum energy which can allow for dS vacua (see for example \cite{Kachru:2003aw, Cicoli:2012fh, Cicoli:2015ylx, Gallego:2017dvd}), the potential for $\tau_1$ can be flat enough to drive inflation. $\tau_1$ plays the r\^ole of the inflaton since, when it is shifted away from its minimum, it is naturally much lighter then the Hubble scale during inflation $H$ whose value is set by the mass of $\tau_1$ close to the minimum, $H^2\simeq W_0^2 /\vo^{10/3}$ (see \cite{Cicoli:2008gp} for more details). 

On the other hand, the other five spectator fields are isocurvature modes which are expected to stay around their minima during inflation. Three of them, $\vo$, $\tau_3$ and $\theta_3$, are heavy fields with a mass larger than $H$, while $\theta_1$ and $\theta_2$ are ultra-light fields since they can develop a non-zero mass only via tiny non-perturbative corrections to the superpotential (\ref{Wsup}). In order to study the possibility of having geometrical destabilisation of any of these entropic directions, we need to focus on the field space metric which looks like:
\be
\mc{L}_{\rm kin}=\frac{\partial^2 K}{\partial T_i \partial \bar{T}_j}\,\partial_\mu T_i \partial^\mu \bar{T}_j
=\frac{\gamma_{ij}}{2}\left(\partial_\mu \tau_i \partial^\mu \tau_j +\partial_\mu \theta_i \partial^\mu \theta_j \right).
\label{Lkin}
\ee
The field space is curved, and so the kinetic terms can be diagonalised exactly only locally. However, in LVS models, we can use the exponentially large overall volume $\vo$ as an excellent expansion parameter to obtain leading order results. Thus if we transform the real parts of the K\"ahler moduli as \cite{Burgess:2010bz}:
\be
\tau_1= e^{\frac{2}{\sqrt{3}}\phi_1+\sqrt{\frac23}\phi_2 +\frac12 \phi_3^2}\,, \qquad\
\vo= e^{\sqrt{\frac32}\phi_2}\,, \qquad
\tau_3= \left(\frac{3}{4\alpha\lambda_3}\right)^{2/3} e^{\sqrt{\frac23}\phi_2} \phi_3^{4/3}\,,
\label{tauTrans}
\ee
the kinetic Lagrangian (\ref{Lkin}) for the real parts simplifies to:
\bea
\mc{L}_{\rm kin}^{(\phi)} &=&\frac12 \partial_\mu \phi_1 \partial^\mu \phi_1  \left(1 + \frac34 \phi_3^2\right) + \frac12 \partial_\mu \phi_2 \partial^\mu \phi_2 \\ \nonumber 
&+& \frac12 \partial_\mu \phi_3 \partial^\mu \phi_3 \left(1 - \frac34  \phi_3^2 + \frac{9}{16}  \phi_3^4 \right) + \frac{3\sqrt{3}}{8} \phi_3^3\,\partial_\mu \phi_1 \partial^\mu \phi_3 \,. \nonumber
\eea
Notice that this expression is diagonal at leading order since (\ref{tauTrans}) implies $\phi_3^2 \sim \mc{O}(\vo^{-1})\ll 1$, while subleading corrections induce a kinetic coupling between the heavy field $\phi_3$ and the canonically normalised inflaton $\phi_1$. However we shall show below that this field is heavy enough to prevent any geometrical destabilisation. Moreover, we point out that in the kinetic Lagrangian (\ref{Lkin}) there is no mixing between real and imaginary parts of the K\"ahler moduli. The kinetic terms for the axions read: 
\bea
\mc{L}_{{\rm kin}}^{(\theta)}&=& \frac{1}{4\tau_1^2}\partial_\mu \theta_1\partial^\mu \theta_1 +\frac{\alpha^2 \lambda_3\tau_3^{3/2}}{2\vo^2\sqrt{\tau_1}} \partial_\mu\theta_1\partial^\mu\theta_2 + \frac{\alpha^2\tau_1}{2\vo^2}\partial_\mu\theta_2\partial^\mu\theta_2 \nonumber \\
&-&\frac{3\alpha\lambda_3\sqrt{\tau_3}}{4\tau_1\vo}\partial_\mu\theta_1\partial^\mu\theta_3-\frac{3\alpha^2\lambda_3\sqrt{\tau_1}\sqrt{\tau_3}}{2\vo^2}\partial_\mu\theta_2\partial^\mu\theta_3+\frac{3\alpha\lambda_3}{8\vo\sqrt{\tau_3}}\partial_\mu\theta_3\partial^\mu\theta_3\,,
\label{LkinAx}
\eea
where $\tau_1$, $\vo$ and $\tau_3$ are given by (\ref{tauTrans}). The kinetic Lagrangian (\ref{LkinAx}) clearly shows that the two ultra-light axions $\theta_1$ and $\theta_2$ are kinetically coupled to the canonically normalised inflaton $\phi_1$. It is therefore crucial to analyse the contribution to the isocurvature power spectrum of each of these two entropic modes. We shall find below that the effective mass-squared of one of these two ultra-light axions is negative during inflation while the other always remains positive. This result justifies the fact that we will study the dynamics of the system by focusing just on the 2-field subspace spanned by the inflaton $\phi_1$ and the unstable isocurvature direction, as summarised in Sec. \ref{sec:GeomDest}. 

We shall find that which of the two axions is unstable depends on the particular realisation of Fibre Inflation. Thus we conclude this section by providing a brief description of two ways to generate the inflationary potential which are qualitatively similar but quantitatively slightly different: 
\bi
\item \textbf{Right-left inflation}

Kaluza-Klein and winding 1-loop open string corrections to $K$ \cite{Berg:2005ja,Berg:2007wt,Cicoli:2007xp} generate a potential for the inflaton shifted from its minimum, i.e. $\phi_1=\langle\phi_1\rangle+\hat\phi_1$, of the form \cite{Cicoli:2008gp}:
\be
V = V_0 \left(3 -4\, e^{-\frac{\hat\phi_1}{\sqrt{3}}}+ e^{-\frac{4\hat\phi_1}{\sqrt{3}}}\right),
\label{Vgs}
\ee
where we included an uplifting term to achieve a dS vacuum after the end of inflation and we neglected additional subleading loop effects which would lift the flatness of the inflationary plateau at very large field values. Notice that this is a case of right-left inflation where $\hat\phi_1$ evolves from positive and large field values to smaller ones towards the end of inflation. Hence $V_{\hat\phi_1}>0$ during inflation.
\newpage
\item \textbf{Left-right inflation} 

The inflationary potential can receive non-negligible contributions not just from string loops but also from higher derivative $\alpha'$ effects which at the level of the $4$-dimensional effective field theory appear as $F^4$ terms \cite{Ciupke:2015msa, Grimm:2017okk}. When these effects are combined with Kaluza-Klein string loops, the inflationary potential looks like \cite{Broy:2015zba}:
\be
V = \tilde{V}_0 \left(1- e^{\frac{1}{\sqrt{3}}\hat\phi_1} \right)^2\,,
\label{eq:simplifiedalphaprime}
\ee
where again we included an uplifting term and we ignored subdominant contributions which would spoil the inflationary plateau for $\hat\phi_1$ negative and very large in absolute value. Contrary to the previous case, this is therefore a realisation of left-right inflation where $V_{\hat\phi_1}<0$ during inflation.
\ei

\subsection{Stability of heavy fields}

The leading order potential (\ref{VLVS}) depends just on the three fields $\phi_2$, $\phi_3$ and $\theta_3$, which therefore turn out to be heavier that the inflaton $\phi_1$. We shall now consider the 2-field subspaces spanned by $\phi_1$ and each of these heavy fields separately, and show that all of them are heavy enough to ensure the absence of any geometrical destabilisation. The field space metric and the Ricci scalar of these 2-dimensional subspaces are listed in Tab. \ref{tab:TabHF}. 

\begin{table}
\begin{tabular}{|M{1.0cm}|M{4.0cm}|M{3.2cm}|M{5cm}|N}
  \hline
  &  $\theta_3$ &  $\phi_2$ &  $\phi_3$& \\[10pt]
  \hline
  $\gamma_{ij}$ &
  $\begin{pmatrix}
  1+\frac34 \phi_3^2  &                         0                         \\ 
  0  & \frac{3\alpha\lambda_3}{4 \vo\sqrt{\tau_3}}
  \end{pmatrix}$ & 
  $\begin{pmatrix}
  1+\frac34\phi_3^2  &      0  \\
  0                                 &      1  \\ 
  \end{pmatrix} $&
  $\begin{pmatrix} 
  1+\frac34 \phi_3^2  &      \frac{3\sqrt{3}}{8}\phi_3^3     \\
  \frac{3\sqrt{3}}{8}\phi_3^3          &   1-\frac34\phi_3^2+\frac{9}{16}\phi_3^4 \\  
  \end{pmatrix}$& \\[10pt]
  \hline
  $R$ & $0$& $0$& $-3/2$&  \\[10pt]
  \hline
\end{tabular}
\caption{Field space metric and Ricci scalar for the 2-field subspaces spanned by the inflaton $\phi_1$ and each of the three heavy fields.}
\label{tab:TabHF}
\end{table}

The simpler cases to analyse involve $\phi_2$ and $\theta_3$ since the metric is diagonal at this level of approximation (perturbative and non-perturbative corrections to the K\"ahler potential will definitely induce subdominant non-diagonal entries), and so the scalar curvature is vanishing. Moreover, we expect the heavy fields to sit around their minima, i.e. $V_{\phi_2} = V_{\theta_3} \simeq 0$, and inflation to be driven by $\phi_1$, i.e. $\alpha_{\phi_1}\simeq 1$ while $\alpha_{\phi_2}\simeq \alpha_{\theta_3}\simeq 0$. Therefore the trajectory is geodesic in both cases (denoting the $2$ heavy fields collectively as $\phi_{\rm h}$):
\be
\eta_\perp\simeq\frac{1}{H\dot{\phi}_0}\left(\alpha_{\phi_1} \frac{V_{\phi_{\rm h}}}{f} - \alpha_{\phi_{\rm h}} V_{\phi_1}\right)\simeq 0 \,.
\ee 
The effective mass-squared \eqref{eq:m2eff} therefore reduces simply to: 
\be
m_{\theta_3,\,{\rm eff}}^2 \simeq \frac{V_{\theta_3\theta_3}}{f^2} \simeq \frac{W_0^2}{\vo^2} \gg m_{\phi_2,\,{\rm eff}}^2 \simeq V_{\phi_2\phi_2} \simeq \frac{W_0^2}{\vo^3} \gg H^2 \simeq \frac{W_0^2}{\vo^{10/3}}\,.
\ee  
Similar considerations apply to the subspace spanned by $\phi_1$ and $\phi_3$ since the field space is flat at leading order. However subleading corrections proportional to $\phi_3^2\sim\mc{O}(\vo^{-1})\ll 1$, induce non-vanishing Christoffel symbols and Ricci scalar:
\be
\Gamma^{\phi_1}_{\phi_3\phi_3} = \frac{9\sqrt{3}}{8} \phi_3^2 \left(1 - \frac12\phi_3^2 + \frac{3}{16}\phi_3^4\right)\sim\mc{O}\left(\frac{1}{\vo}\right)\,,\qquad R = -\frac32\,.
\ee
The effective mass-squared of the heavy field $\phi_3$ for $V_{\phi_3}\simeq \alpha_{\phi_3} \simeq 0$ and $\alpha_{\phi_1}\simeq 1$, which imply $\eta_\perp\simeq 0$, reduces to: 
\be
m_{\phi_3,\,{\rm eff}}^2 \simeq V_{\phi_3\phi_3} - \Gamma^{\phi_1}_{\phi_3\phi_3} V_{\phi_1}-\frac32\epsilon H^2\,.
\ee
This quantity is clearly positive regardless of the shape of the inflationary potential since:
\be
V_{\phi_3\phi_3} \simeq \frac{W_0^2}{\vo^2}\gg \left\{\begin{array}{ll}
\Gamma^{\phi_1}_{\phi_3\phi_3} V_{\phi_1} \simeq \frac{W_0^2 \sqrt{\epsilon}}{\vo^{13/3}} \\[5pt]
\frac32\epsilon H^2 \simeq \frac{W_0^2\epsilon}{\vo^{10/3}} \\
\end{array}\right. .
\ee
We have therefore shown that, as expected, all heavy fields remain stable during Fibre Inflation.

\subsection{Potential destabilisation of ultra-light axions}

In this section we analyse the behaviour of the two ultra-light axionic modes $\theta_1$ and $\theta_2$. The metric of the $2$-dimensional field spaces spanned by the inflaton $\phi_1$ and either $\theta_1$ or $\theta_2$ takes the same form as (\ref{gammaij}) if we neglect subdominant $\phi_3$-dependent corrections. Notice that the kinetic function $f(\phi_1)$ becomes $\phi_1$-dependent after (\ref{tauTrans}) is used to express $\tau_1$ in terms of the canonically normalised fields $\phi_2$ and $\phi_3$ which are fixed at their minima. Given that $f(\phi_1)$ is a particular case of the more general form (\ref{fform}), the scalar curvature is constant and negative. These geometrical quantities are summarised in Tab. \ref{Tab:ULFFIGD} where the quantities $A_+$ and $A_-$ depend on the background values of the heavy fields.

\begin{table}
\begin{center}
\begin{tabular}{|M{1.0cm}|M{4.0cm}|M{4.0cm}|N}
  \hline
  &  $\theta_1$ &  $\theta_2$ & \\[10pt]
  \hline
  $\gamma_{ij}$ &
  $\begin{pmatrix}
  1  &                         0                         \\ 
  0  & A_-^2\, e^{-\frac{4}{\sqrt{3}}\phi_1}
  \end{pmatrix}$ & 
  $\begin{pmatrix}
  1  &      0  \\
  0                                 &      A_+^2\,e^{\frac{2}{\sqrt{3}}\phi_1}  \\ 
  \end{pmatrix} $ & \\[10pt]
  \hline
  $R$ & $-8/3$& $-2/3$&  \\[10pt]
  \hline
\end{tabular}
\caption{Field space metric and Ricci scalar for the 2-field subspaces spanned by the inflaton $\phi_1$ and each of the two ultra-light axions.}
\label{Tab:ULFFIGD}
\end{center}
\end{table}

In the case where $\theta_1$ and $\theta_2$ are exactly massless, we find that one of them is always unstable. In order to solve this potential issue, we investigate the possibility of stopping  the exponential growth of the corresponding isocurvature perturbations by turning on a tiny but non-zero mass for this entropic mode. Let us therefore study these two different cases separately.

\subsubsection{Massless case}
\label{MasslessCase}

The analysis of the possible geometrical destabilisation of $\theta_1$ and $\theta_2$ can be borrowed from Sec. \ref{PotDest} where we already discussed the case where the spectator fields are massless and $\phi_1$ drives inflation in a single-field approximation. Using the result (\ref{eq:meffULFfExpSimpl}) under the condition (\ref{Import}) we therefore conclude that one of the two axions is always stable while the perturbations of the other experience a geometrical instability. In particular, it is the sign of $V_{\hat\phi_1}$ that determines which of the two axions is unstable. For $V_{\hat\phi_1}>0$, as in the case of right-left inflation, $\theta_1$ is unstable while $\theta_2$ is stable. On the contrary, for $V_{\hat\phi_1}<0$, as in the case of left-right inflation, $\theta_1$ is stable while $\theta_2$ becomes unstable. 

These results have been obtained analytically in the single-field approximation where $\alpha_{\phi_1}\simeq 1$ and $\alpha_{\theta_i}\simeq 0$ with $i=1,2$. However they hold more generically as we will show now via a more general semi-analytic study and a detailed numerical analysis. 

As pointed out above, the metric has the same form as (\ref{gammaij}) with $f = f_0\,e^{-k_1 \hat\phi_1}$, where $f_0 = A_-\,e^{-k_1 \langle\phi_1\rangle}$ and $k_1=2/\sqrt{3}$ for $\theta_1$, while $f_0 = A_+\,e^{-k_1 \langle\phi_1\rangle}$ and $k_1=-1/\sqrt{3}$ for $\theta_2$. The equations of motion which govern the evolution of the system are very similar to the ones studied in Sec. \ref{EoM} for the case of a quintessence-like potential. We shall therefore use the same results, translating them for the case of Fibre Inflation. In particular, the second equation in \eqref{eq:EOMSstandardphi1phi2} does not depend on the inflationary potential, and so it applies exactly also to our case after identifying $\phi_2$ with either $\theta_1$ or $\theta_2$. Its solution is given by (\ref{eq:standardfphi2}) which in our case becomes:
\be
\left(f\theta_i'\right)(N) = \left(f\theta_i'\right)(0) \,e^{-\lambda(N)}\,,\qquad \forall i=1,2 \,,
\label{res}
\ee
where for $\epsilon\ll 1$ the exponent $\lambda(N)$ can be approximated as:
\be
\lambda(N) \simeq 3N - k_1 \left(\hat\phi_1(N)-\hat\phi_1(0)\right)\,.
\label{exponent}
\ee
The functional dependence of the inflaton $\phi_1$ on the number of e-foldings $N$ depends on the particular form of the inflationary potential. Let us therefore consider separately the case of right-left \cite{Cicoli:2008gp} and left-right inflation \cite{Broy:2015zba}.

\bi
\item \textbf{Right-left inflation}

For right-left inflation the scalar potential is given by (\ref{Vgs}) which in the inflationary plateau region can be  very well approximated as:
\be
V \simeq V_0 \left(3-4\,e^{-k_2\hat\phi_1}\right),\qquad\text{with}\quad k_2 = \frac{1}{\sqrt{3}}\,.
\label{Vappr}
\ee
The number of e-foldings $N$ in the single-field slow-roll approximation is given by:
\be
N (\hat\phi_1) =\int_{\hat\phi_1}^{\hat\phi_1(0)} \frac{V}{V_{\hat\phi_1}}\, d\hat\phi_1 
= \frac94 \left(e^{k_2 \hat\phi_1(0)}-e^{k_2\hat\phi_1}\right)-\sqrt{3}\left(\hat\phi_1(0)-\hat\phi_1\right)\,.
\ee
This expression cannot be inverted exactly but we can still express the inflaton at leading order as \cite{Cicoli:2018cgu}:
\be
\hat\phi_1(N)- \hat\phi_1(0) \simeq \frac{1}{k_2} \ln\left(1-\frac{4 N}{9}\,e^{-k_2\hat\phi_1(0)}\right)\,,
\ee
where $\hat\phi_1(0)$ corresponds to the value of the inflaton at CMB horizon exit. It is easy to see that $50-60$ e-foldings of inflation correspond to $\hat\phi_1(0)\sim \mc{O}(6)$. Substituting this result into the solution for the velocity of the ultra-light axions (\ref{res}), we find that the exponent (\ref{exponent}) scales with the number of e-foldings as:
\be
\lambda(N) = 3 N - \frac{k_1}{k_2} \ln\left(1-\frac{4 N}{9}\,e^{-k_2\hat\phi_1(0)}\right)\,.
\ee
This quantity is always positive for both $k_1=2/\sqrt{3}$ and $k_1=-1/\sqrt{3}$, implying that, regardless of the initial conditions, the velocity of the ultra-light axions goes very quickly to zero, and so the system relaxes rapidly to the simple case studied above with $\eta_\perp=0$, $\alpha_{\phi_1}\simeq 1$ and $\alpha_{\theta_i}\simeq 0$ with $i=1,2$. 

\item \textbf{Left-right inflation}

In the case of left-right inflation, the number of e-foldings derived from the inflationary potential \eqref{eq:simplifiedalphaprime} in the slow-roll approximation is given by: 
\be
N (\hat\phi_1) =\frac{1}{2 k_2}\int_{\hat\phi_1(0)}^{\hat\phi_1}\left( e^{-k_2\hat\phi_1} - 1\right)d\hat\phi_1
= \frac{1}{2 k_2^2} \left(e^{-k_2\hat\phi_1(0)}-e^{-k_2\hat\phi_1(0)}\right)-\frac{1}{k_2}(\hat\phi_1-\hat\phi_1(0)) \,.
\ee
Again, even if it is not possible to invert this expression exactly, we can still obtain the following leading order approximation for the inflaton field:
\be
\hat\phi_1(N) - \hat\phi_1 (0) = -\frac{1}{k_2} \ln\left(1-\frac{2 N}{3}\, e^{k_2\hat\phi_1(0)}\right)\,,
\label{res1}
\ee 
where $\hat\phi_1(0)<0$ since inflation proceeds from left to right starting from inflaton values which are negative and large in absolute value. If the result (\ref{res1}) is substituted into the expression (\ref{exponent}) for the exponent of the solution (\ref{res}) for the velocity of the isocurvature modes, we find:
\be
\lambda(N) = 3 N + \frac{k_1}{k_2} \ln\left(1-\frac{2 N}{3}\, e^{k_2\hat\phi_1(0)}\right)\,.
\ee
It is easy to realise that this quantity is again always positive for both $k_1=2/\sqrt{3}$ and $k_1=-1/\sqrt{3}$. Hence also in this case, regardless of the initial conditions, the system approaches very rapidly a geodesic trajectory with $\eta_\perp=0$.
\ei

\begin{figure}[!h]
  \centering
  \begin{minipage}[b]{0.29\textwidth}
    \includegraphics[width=\textwidth]{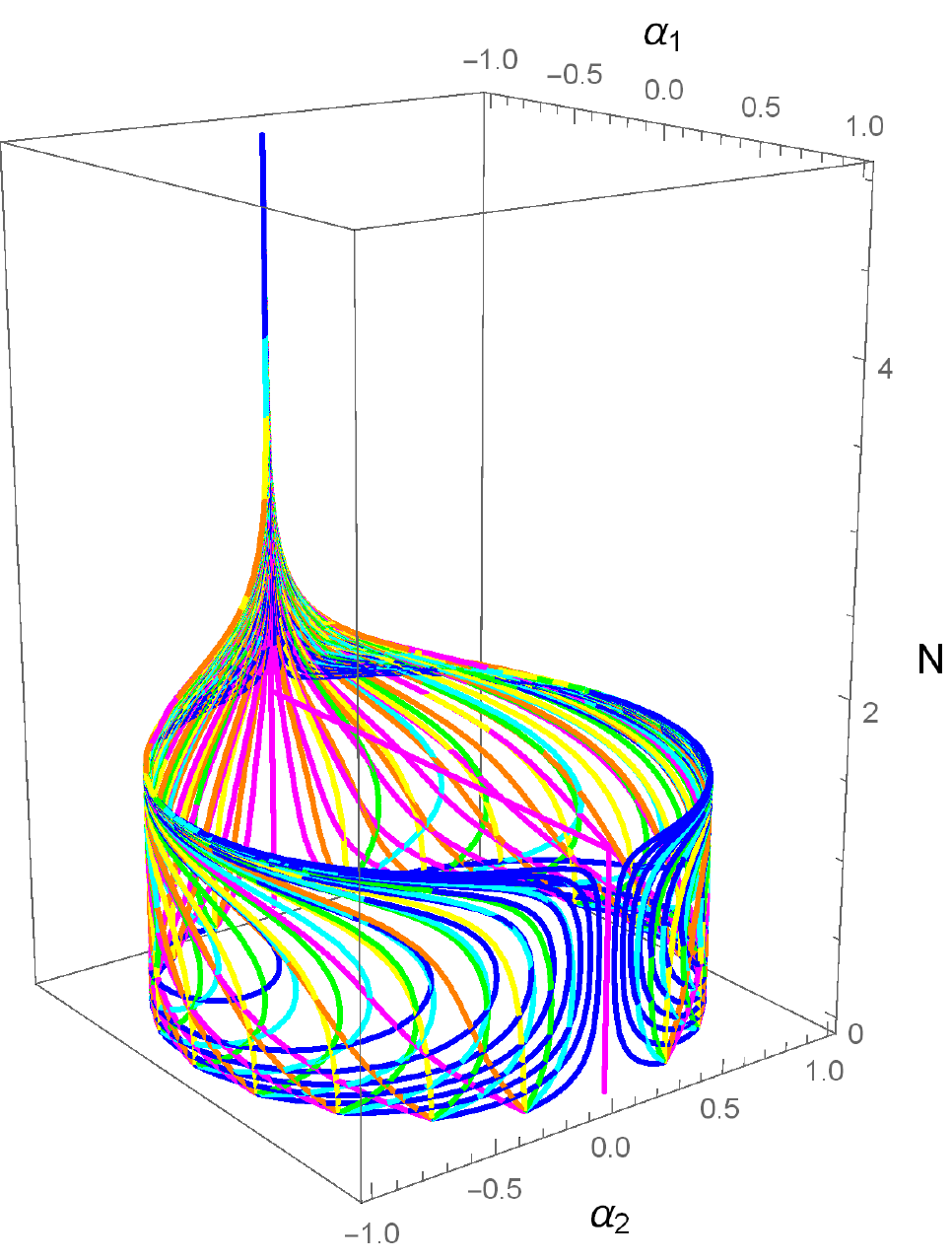}
  \end{minipage}
  \hfill
  \begin{minipage}[b]{0.29\textwidth}
    \includegraphics[width=\textwidth]{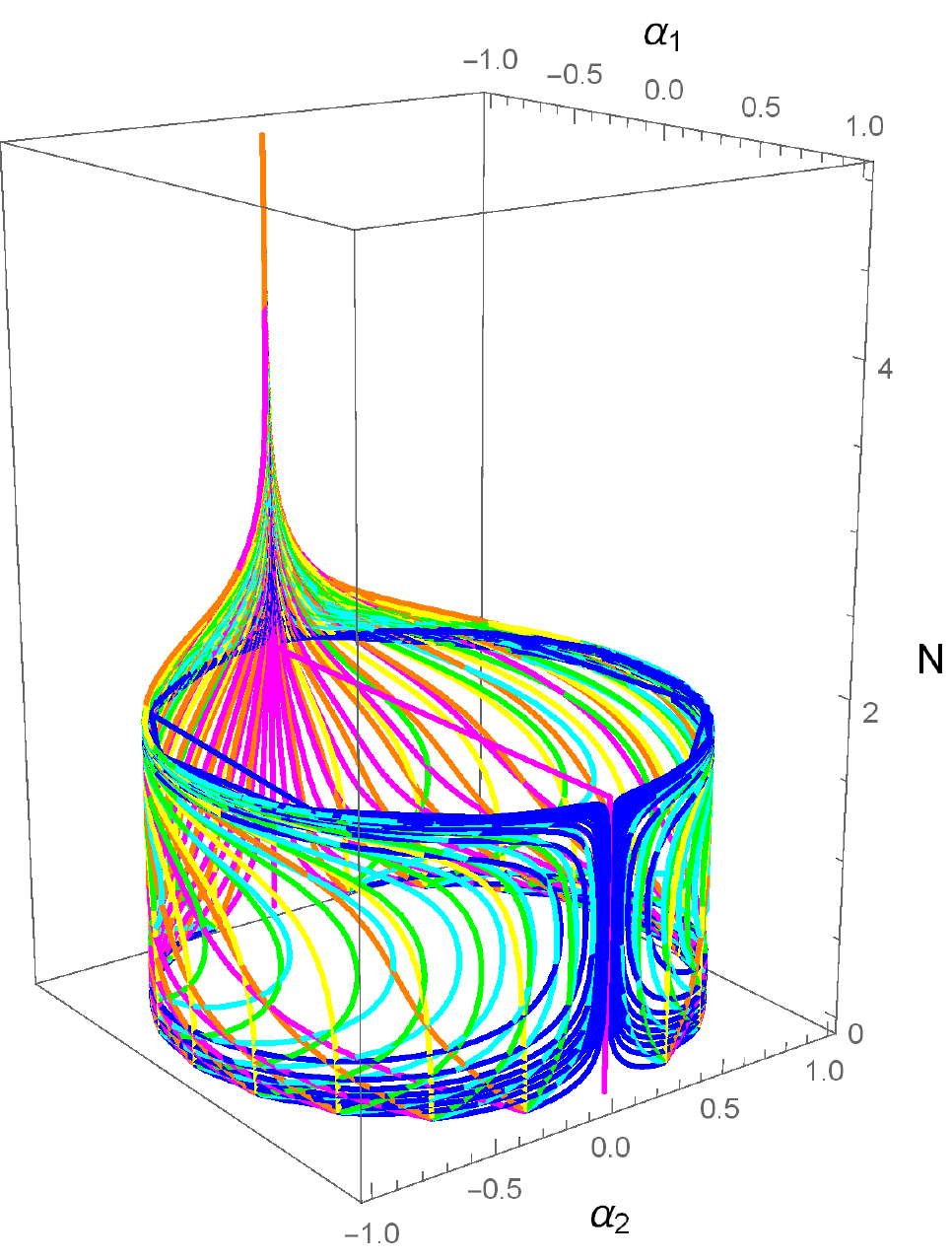}
  \end{minipage}\hfill
  \begin{minipage}[b]{0.36\textwidth}
    \includegraphics[width=\textwidth]{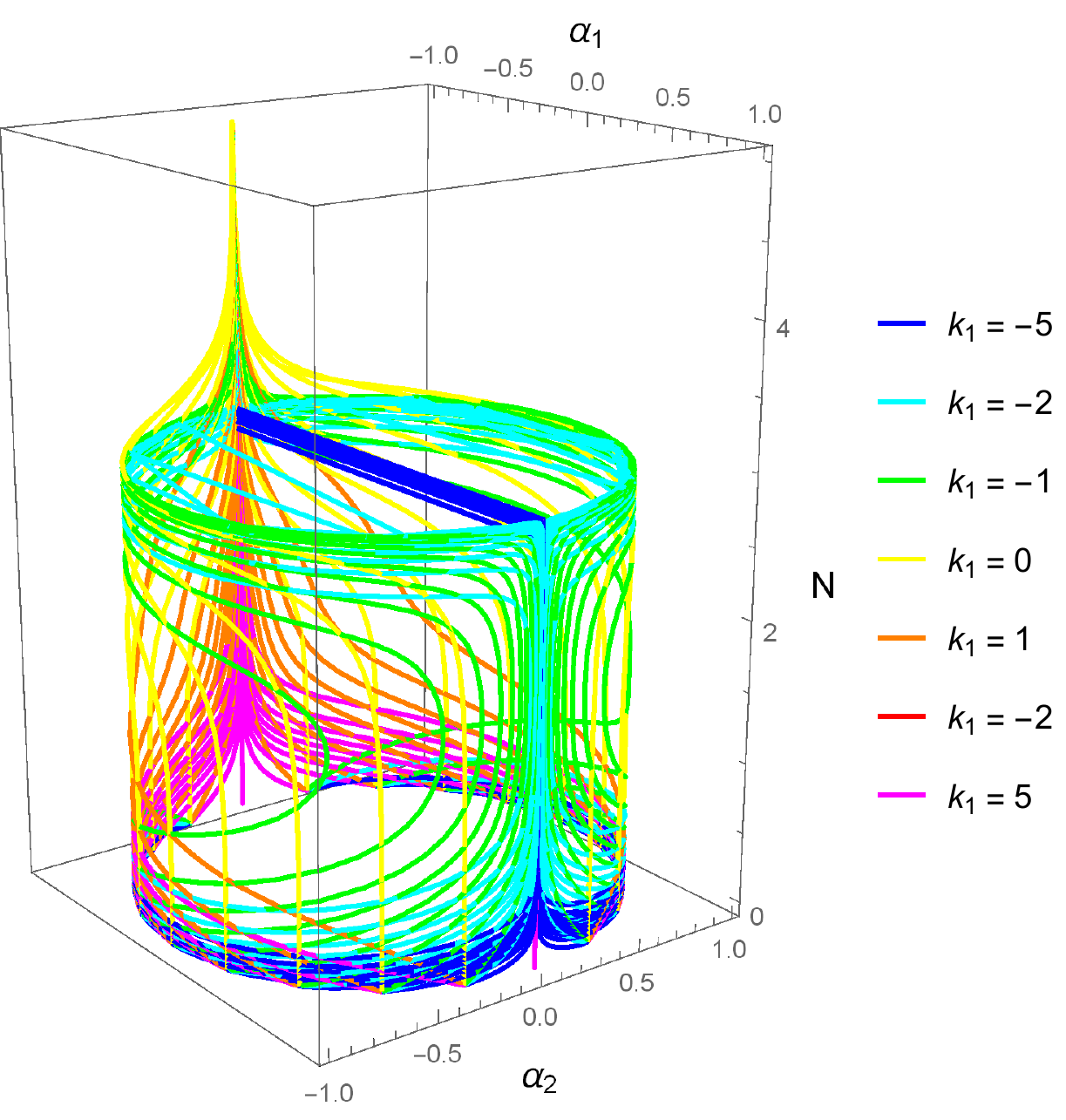}
  \end{minipage}
  \caption{Evolution of the system for several values of $k_1$ and different initial kinetic energies, $\epsilon(0)=1$ (left), $\epsilon(0)=2$ (centre), $\epsilon(0)=2.99$ (right).}
  \label{Fig5}
\end{figure}

We have checked these conclusions by performing a full numerical solution of the equations of motion governing the evolution of the system for both right-left and left-right inflation. We present now the numerical results just for right-left inflation since they are qualitatively very similar in the case of left-right inflation. Without loss of generality we considered $f_0=V_0=1$, $\hat\phi_1(0)=5.8$ and $\theta_i(0)=0$ for $i=1,2$. Fig. \ref{Fig5} shows clearly that for different values of the initial kinetic energy and for several exponents of the kinetic function, $k_1=\{-5,-2,-1,0,1,2,5\}$, the system always evolves  towards a single-field behaviour. 

\begin{figure}[!h]
  \begin{minipage}[b]{0.57\textwidth}
    \includegraphics[width=\textwidth]{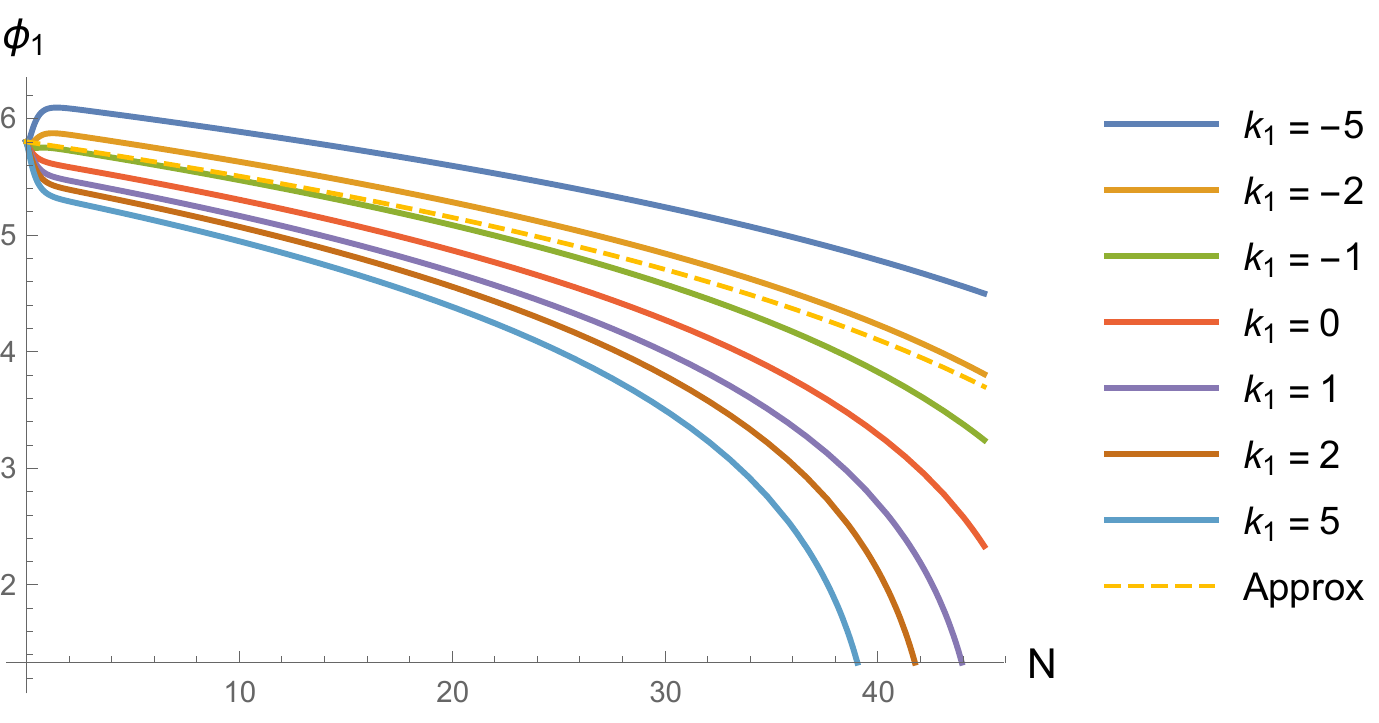}
  \end{minipage}
  \hfill
  \begin{minipage}[b]{0.42\textwidth}
    \includegraphics[width=\textwidth]{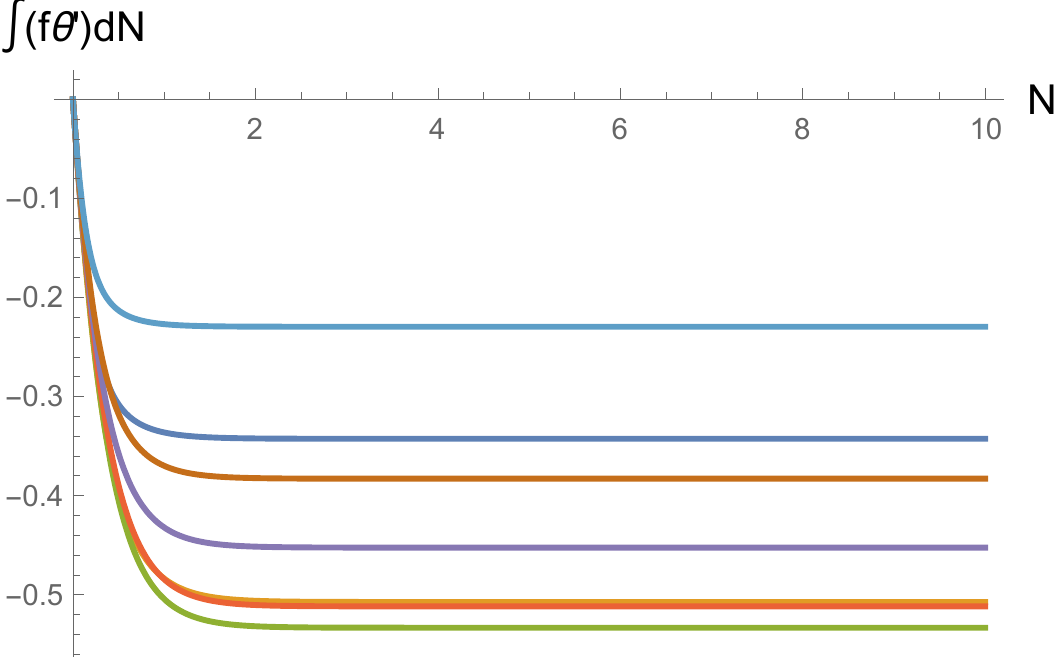}
  \end{minipage}
  \caption{Trajectories of the physical fields $\hat\phi_1(N)$ and $\int_0^N f(\phi_1(\tilde{N}))\theta_i'(\tilde{N}) d\tilde{N}$ for different values of $k_1$ and $\epsilon(0)=1$, $\hat\phi_1'(0) = \sqrt{2} \,\cos\left(\omega\right)$, $\left(f\theta_i'\right)(0) = \sqrt{2} \,\sin\left(\omega\right)$ with $\omega=7\pi/5$. The dashed line represents the single-field analytical approximation with zero initial velocity.}
  \label{fig:phi1fphi2differentk1epsilon1FI}
\end{figure}

\begin{figure}[!h]
  \centering
  \begin{minipage}[b]{0.55\textwidth}
    \includegraphics[width=\textwidth]{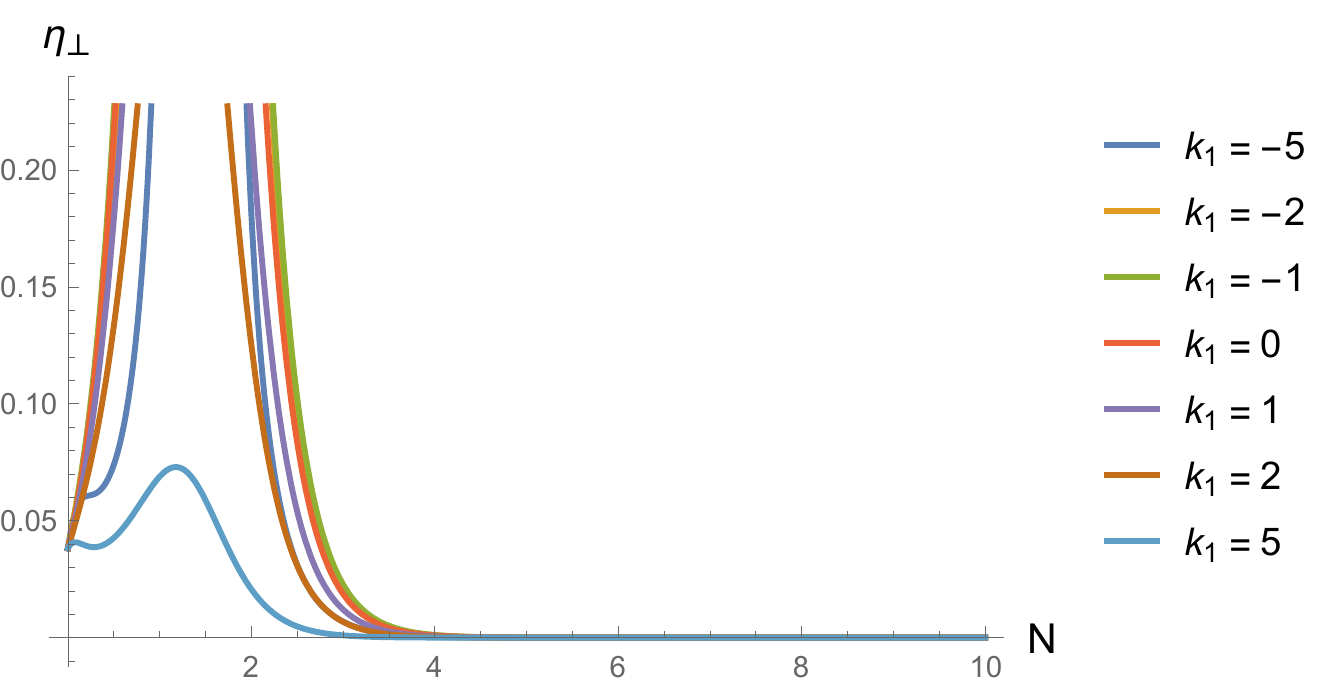}
  \end{minipage}
  \hfill
  \begin{minipage}[b]{0.44\textwidth}
    \includegraphics[width=\textwidth]{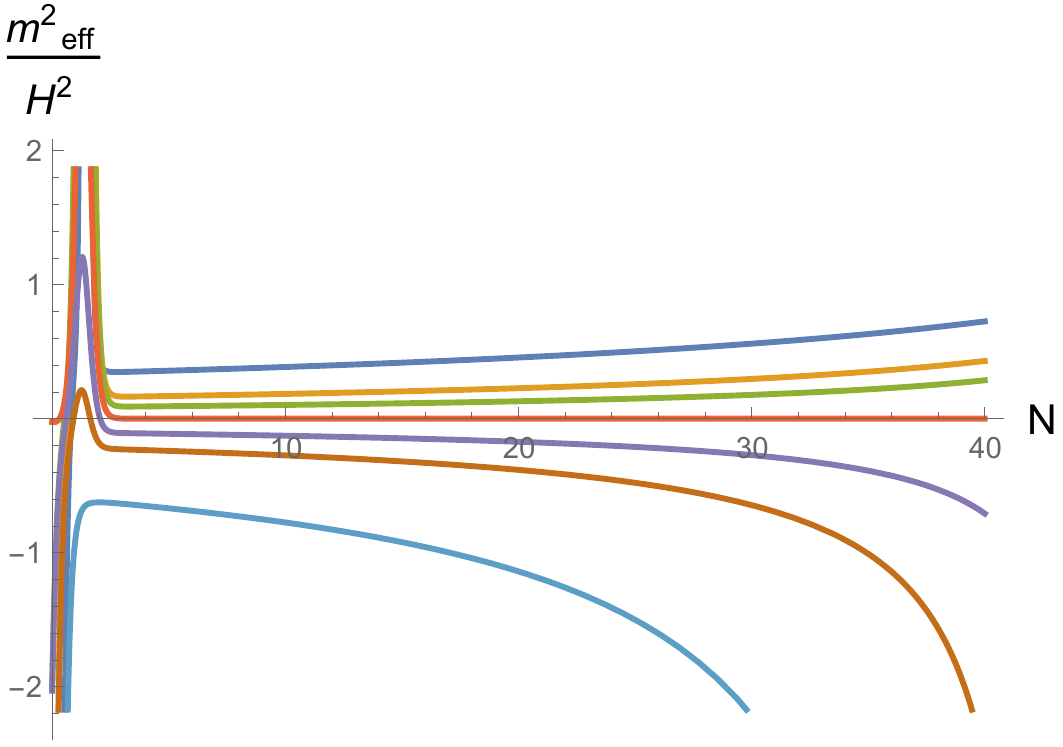}
  \end{minipage}
  \caption{$\eta_\perp$ (left) and $m_{\perp,\,{\rm eff}}^2/H^2$ (right) as a function of the number of e-foldings for different values of $k_1$, setting again $\epsilon(0)=1$, $\hat\phi_1'(0) = \sqrt{2} \,\cos\left(\omega\right)$ and $\left(f\theta_i'\right)(0) = \sqrt{2} \,\sin\left(\omega\right)$ with $\omega=7\pi/5$.}
  \label{fig:m2effetaperpdifferentk1epsilon1FI}
\end{figure}

Fig. \ref{fig:phi1fphi2differentk1epsilon1FI} presents instead the trajectory of the physical fields $\hat\phi_1(N)$ and $\int_0^N (f\theta_i')(\tilde{N}) d\tilde{N}$ for different values of $k_1$. Finally Fig. \ref{fig:m2effetaperpdifferentk1epsilon1FI} shows that any value of $k_1$ leads to a geodesic motion with $\eta_\perp=0$ but the effective mass-squared of the isocurvature perturbations can remain positive only for $k_1<0$, implying that $\theta_1$ (with $k_1=2/\sqrt{3}$) is unstable, while $\theta_2$ (with $k_1=-1/\sqrt{3}$) does not experience any geometrical destabilisation. Notice that the situation is reversed in the case of left-right inflation.

\subsubsection{Massive case}
\label{MassiveCase}

In Sec. \ref{MasslessCase} we have shown that in Fibre Inflation models, when the axions are considered as exactly massless, one of them always experiences  geometrical destabilisation. However in a full quantum model, these entropic modes are expected to receive a tiny but non-zero mass from non-perturbative corrections to the superpotential (\ref{Wsup}) which break their perturbative shift symmetry. Let us investigate now if these non-perturbative effects can be large enough to avoid any geometrical destabilisation problem and, at the same time, small enough to prevent any modification of the inflationary dynamics.

As we have seen in the analysis of the massless case, the single-field approximation with $\alpha_{\phi_1}\simeq 1$ and $\alpha_{\theta_i}\simeq 0$ with $i=1,2$ provides a very good description of the more general dynamics of the system. Hence the equation to analyse is \eqref{eq:m2effalpha} which after identifying $\phi_2$ with $\theta_i$ and setting $\alpha_{\phi_1}\simeq 1$ and $\alpha_{\theta_i}\simeq 0$ for $i=1,2$, takes the form:
\be
m_{\theta_i,\,{\rm eff}}^2 = \left(\frac{V_{\theta_i\theta_i}}{f^2}+\frac{f_{\hat\phi_1}}{f} V_{\hat\phi_1}+3\frac{V_{\theta_i}^2}{\dot{\phi}_0^2 f^2}\right)-\dot{\phi}_0^2\frac{f_{\hat\phi_1\hat\phi_1}}{f}\,.
\label{eqnew}
\ee
If we write the kinetic function as in (\ref{fform}) and we recall the slow-roll condition (\ref{Import}), the effective mass-squared (\ref{eqnew}) for the dangerous entropic modes simplifies to:
\be
m_{\theta_i,\,{\rm eff}}^2 \simeq - \lambda \left|V_{\hat\phi_1}\right| \left[1 - \frac{1}{\lambda f^2 \sqrt{2\epsilon}}\left(\frac{V_{\theta_i\theta_i}}{V} +\frac{9}{2\epsilon}\frac{V_{\theta_i}^2}{V^2}\right)\right]\,,
\label{Eqnew}
\ee
where we have used the slow-roll approximations $\dot{\phi}_0^2 = 2\epsilon H^2 \simeq 2\epsilon V/3$ and $\sqrt{2\epsilon} \simeq \left|V_{\hat\phi_1}\right|/V$.

The potential for the entropic directions is generated by $T_i$-dependent non-perturbative corrections to the superpotential (\ref{Wsup}):
\be
W=W_0+A_3\, e^{-a_3 T_3}+A_i\, e^{-a_i T_i}\,,\qquad i=1,2\,,
\label{eq:npcorrectionsW}
\ee
which induce a non-zero potential for the ultra-light axions $\theta_i$ of the form:
\be
V(\theta_i) = \Lambda_i \cos(a_i \theta_i)\,,\qquad\text{with}\quad \Lambda_i = \frac{4 a_i A_i W_0 \tau_i}{\vo^2}\, e^{-a_i\tau_i}\,,\qquad i=1,2\,.
\label{Vtheta}
\ee
Hence we obtain:
\be
V_{\theta_i}^2 = a_i^2\, \Lambda_i^2 \sin^2 (a_i \theta_i)\qquad\text{and}\qquad V_{\theta_i\theta_i} = -a_i^2 \,\Lambda_i \cos(a_i \theta_i)\,.
\ee
The effective mass-squared of the isocurvature perturbations (\ref{Eqnew}) can therefore be rewritten as:
\be
m_{\theta_i,\,{\rm eff}}^2 \simeq - \lambda \left|V_{\hat\phi_1}\right| \left[1 -\frac{a_i^2}{\lambda f^2 \sqrt{2\epsilon}}\left(\frac{9 \delta^2}{2\epsilon} \sin^2 (a_i \theta_i)-\delta \cos(a_i \theta_i) \right)\right]\,,
\label{Eqnew2}
\ee
where $\delta$ is the ratio between the size of the axion potential and the inflationary potential:
\be
\delta \equiv \frac{\Lambda_i}{V(\hat\phi_1)}\,.
\ee
Let us point out that, once $\tau_i$ with $i=1,2$ is written in terms of canonically normalised fields using (\ref{tauTrans}), the axion potential (\ref{Vtheta}) clearly depends on the inflaton $\hat\phi_1$ since:
\be
\Lambda_i=\Lambda_i^{(0)}\, e^{-g_i(\hat\phi_1)}\,,\qquad\text{with}\quad 
\Lambda_i^{(0)} = \frac{4 a_i A_i W_0 \langle\tau_i\rangle}{\vo^2}\,\quad \forall i=1,2\,,
\ee
where:
\be
g_1 (\hat\phi_1) = -2k_2\hat\phi_1+ a_1\langle\tau_1\rangle\, e^{2k_2\hat\phi_1}\,, \qquad
g_2 (\hat\phi_1) = k_2\hat\phi_1 + a_2 \langle\tau_2\rangle\,e^{-k_2\hat\phi_1}\,.
\label{g}
\ee
Notice that for the case of right-left inflation, the dangerous axionic mode is $\theta_1$ and during inflation $\hat\phi_1$ evolves from positive large values to smaller one. On the other hand, in left-right inflation, we need to focus on $\theta_2$ and at the beginning of inflation $\hat\phi_1$ is negative and large in absolute value. Thus in both cases, the axion potential experiences a double exponential suppression, being larger close to the end of inflation and extremely suppressed in the region around CMB horizon exit. 

The fact that the axion potential is $\hat\phi_1$-dependent implies that we cannot make the mass of $\theta_i$ as large as we would like by tuning the underlying parameters $A_i$ and $a_i$ since at a certain point the potential (\ref{Vtheta}) will become of the same order of magnitude of the inflationary potential. This will induce $\mc{O}(1)$ corrections to the inflationary dynamics which would destroy Fibre Inflation as we know it. Hence for consistency we need to impose $\delta \ll 1$, which implies that the two terms proportional to $\delta$ in (\ref{Eqnew2}) are subdominant. 

\begin{figure}[!h]
  \centering
  \begin{minipage}[b]{0.35\textwidth}
    \includegraphics[width=\textwidth]{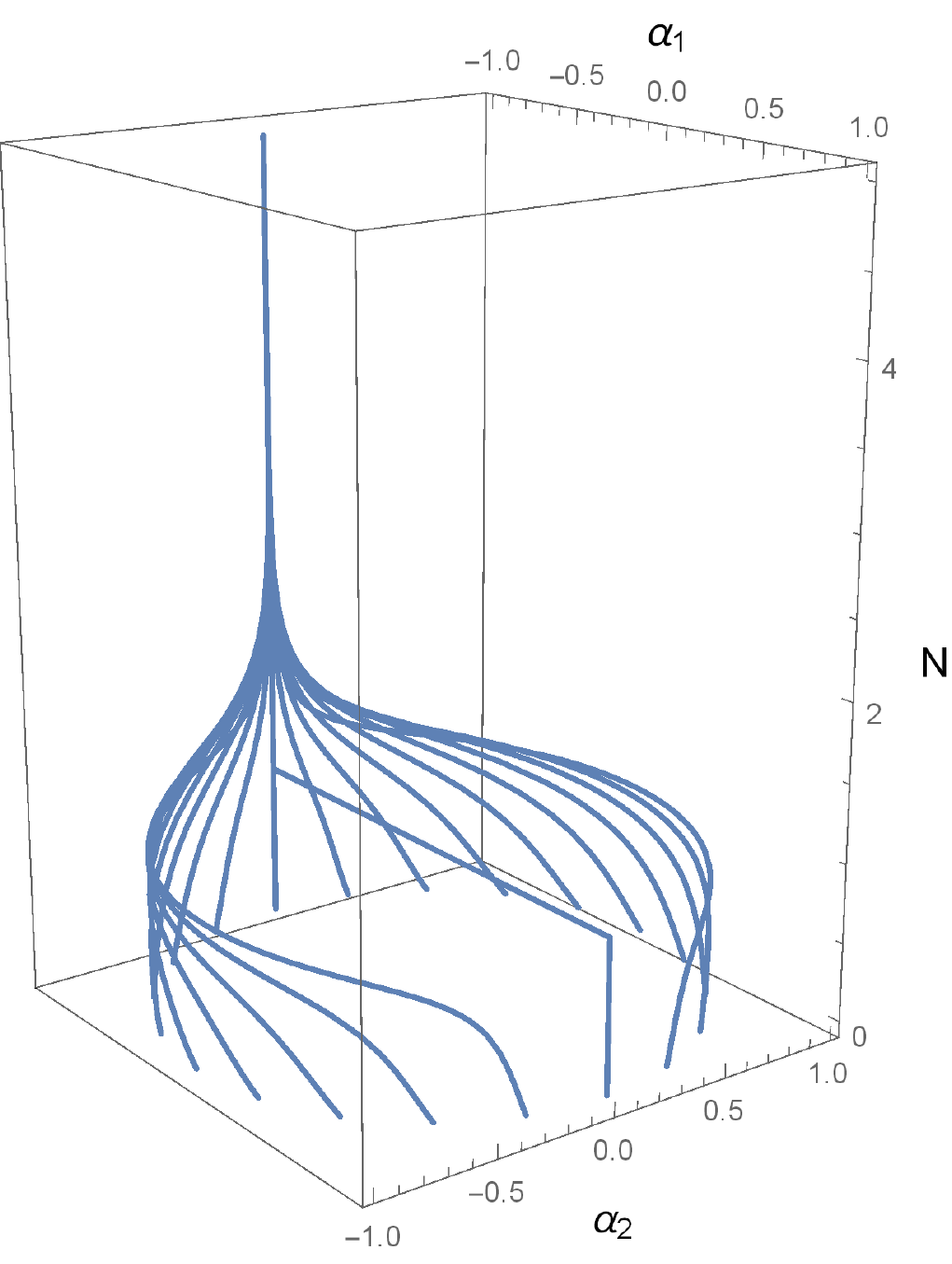}
  \end{minipage}
  \hfill
  \begin{minipage}[b]{0.5\textwidth}
    \includegraphics[width=\textwidth]{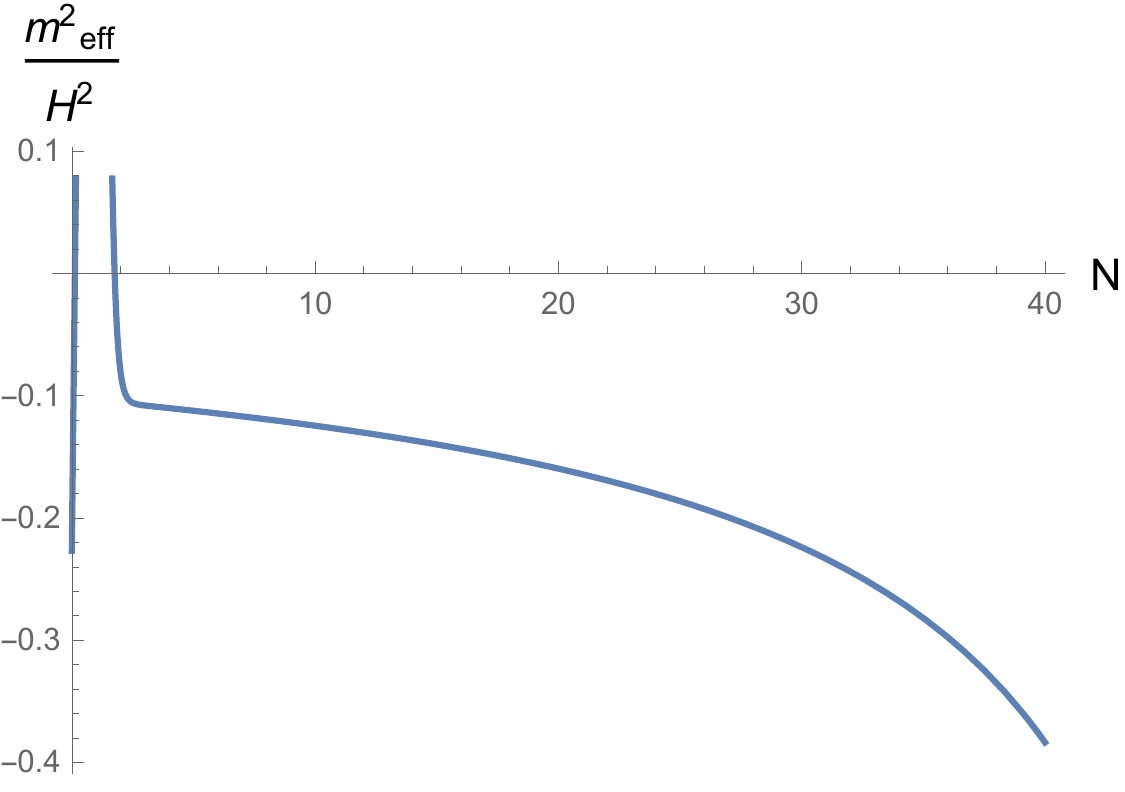}\vspace{20pt}
  \end{minipage}
  \caption{The plot on the left hand side shows the evolution of the $2$-field system of right-left inflation for different initial velocities, $\hat\phi_1(0)=5.8$, $a_1\theta_1(0)=1$, initial kinetic energy $\epsilon(0)=0.1$, $A_1=W_0=1$, $a_1=2\pi$, $\langle\tau_1\rangle=5.43$, $V_0=3.5\times 10^{-11}$ and $\vo=1.8 \times 10^3$. The plot on the right hand side exhibits instead the behaviour of the effective mass-squared of $\theta_1$ for $\hat\phi_1'(0) = \sqrt{2} \,\cos\left(\omega\right)$ and $\left(f\theta_1'\right)(0) = \sqrt{2} \,\sin\left(\omega\right)$ with $\omega=7\pi/5$.}
  \label{fig:alphasandm2effFI}
\end{figure}

Let us stress that, even if $\delta \ll 1$, one of these two terms might actually be the dominant contribution since $f\ll 1$ and $\epsilon\ll 1$, but this can occur only locally around a particular region in field space. In fact, as can be seen from (\ref{g}), $\Lambda_i$ has a double exponential suppression, and so small deviations of the inflaton $\hat\phi_1$ would immediately suppress these positive contributions to $m_{\theta_i,\,{\rm eff}}^2$. We conclude that, even in the presence of non-vanishing scalar potential contributions, the isocurvature fluctuations associated to one of the two ultra-light axions in Fibre Inflation experience an exponential growth, regardless of the particular microscopic realisation of the inflationary model. 

\begin{figure}[!h]
  \centering
    \includegraphics[width=0.6\textwidth]{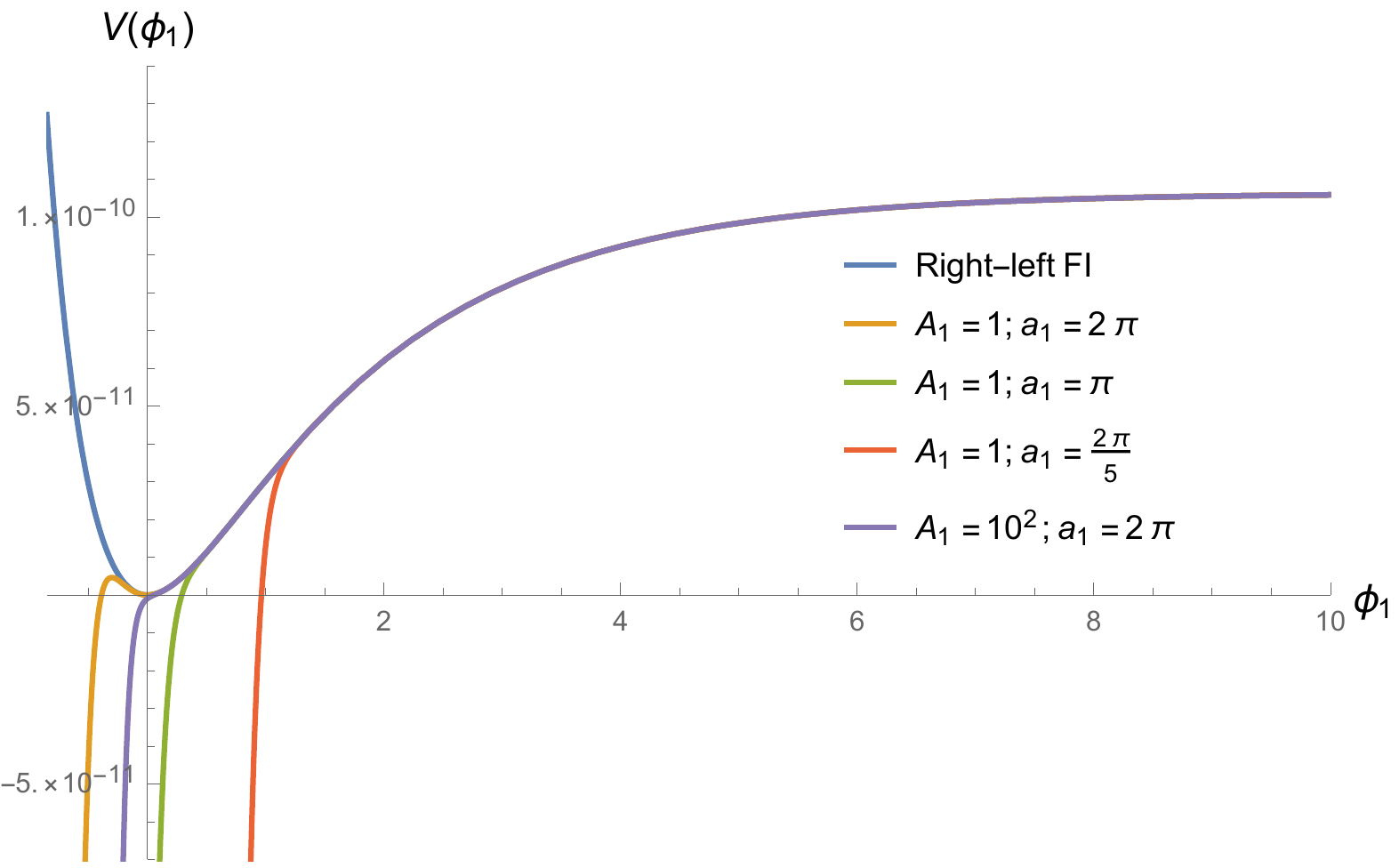}
    \caption{Comparison between the standard single-field and the $2$-field version of right-left inflation for different values of the parameters $A_1$ and $a_1$. Notice that the axionic potential is subleading with respect to the inflationary potential until the end of inflation only for $A_1=1$ and $a_1=2\pi$.}
  \label{fig:FIpotentialhierarchy}
\end{figure}

We checked the validity of these analytic results by performing a full numerical solution of the evolution of the system in the presence of non-perturbative corrections of the form (\ref{eq:npcorrectionsW}). The results for right-left inflation are shown in Fig. \ref{fig:alphasandm2effFI}-\ref{fig:FIpotentialhierarchy}. In particular, Fig. \ref{fig:alphasandm2effFI} shows that the system quickly converges towards a geodesic trajectory and that the effective mass-squared of $\theta_1$ is initially positive due to an appropriate choice of initial conditions but then rapidly settles down to negative values. On the other hand, in Fig. \ref{fig:FIpotentialhierarchy} we see that natural choices of the underlying parameters can keep the axion potential always subleading with respect to the inflationary potential. In this way, the inflationary dynamics is guaranteed to reproduce the one of Fibre Inflation but one of the axionic modes experiences a potential geometrical destabilisation. We finally point out that we obtained numerical results also for left-right inflation and they turn out to be qualitatively very similar.

\section{Conclusions}

The existence of large numbers of massless scalars, the moduli, is a hallmark of string compactification models. For the phenomenological viability of such models it is imperative to generate a mass for the moduli fields, a research area that has seen significant progress over the last 15 years and that often involves the addition of subleading corrections, both perturbative and non-perturbative, to the effective action. Despite the sophistication of current constructions, one can sometimes end up with some remaining massless fields. These are the focus of this work, in particular their r\^ole during inflation.

It has been noted in \cite{Cicoli:2018ccr} that in negatively curved field spaces, massless scalar fields induce an instability at the level of the isocurvature perturbations. This instability arises precisely due to the field space curvature contribution to the effective mass-squared of these perturbations. Whenever the mass-squared becomes negative, one is faced with an uncontrolled growth of the isocurvature 2-point function. 

After gaining some intuition from analysing the simple case of exponential quintessence-like potentials, we studied this instability in the context of Fibre Inflation \cite{Cicoli:2008gp, Burgess:2016owb}, a type IIB string inflation model where the inflationary potential is generated by perturbative corrections to the K\"ahler potential. In this setup there are two axionic fields that remain massless after moduli stabilisation, $\theta_1$ and $\theta_2$, both of which are kinetically coupled to the inflaton. We showed that one of these fields always induces a geometrical instability. For right-left Fibre Inflation models \cite{Cicoli:2008gp}, it is the fibre axion $\theta_1$ that leads to unstable isocurvature perturbations, while in left-right realisations of Fibre Inflation \cite{Broy:2015zba}, the instability is triggered by the base axion $\theta_2$. In both cases we have tried to avoid the instability by giving mass to the axions. We found that although a potential for these fields can be generated by non-perturbative effects, it is not possible to avoid the instability without significantly modifying the dynamics of Fibre Inflation. Furthermore we have numerically probed the system and have shown that this behaviour is independent of the choice of initial conditions. 

At this point, two independent lines of future research arise. On the one hand, one could avoid any potential issue with geometrical destabilisation by making the axions heavy enough. However this would lead to a completely different inflationary model with a truly multi-field dynamics that should be carefully analysed. On the other hand, if one wants to preserve the typical leading order Fibre Inflation dynamics, the exponential growth of the dangerous axionic entropic mode cannot be ignored. Thus in order to determine the ultimate fate of the system one should resort to more sophisticated methods beyond perturbation theory like numerical relativity or the stochastic inflation formalism. Let us simply mention here that, once the isocurvature instability sets in, it is highly possible that the backreaction of the perturbations on the background becomes important and stops the geometrical instability by inducing a turn in the trajectory. 

Let us finally stress that in the analysis of Fibre Inflation with more than one ultra-light entropic direction, in order to make use of the results in the literature concerning the effective mass-squared of the isocurvature modes, we have reduced the field space to two dimensions by projecting out one of the ultra-light directions at a time. Given that only one of the entropic directions turns out to be unstable, this seems reasonable though it would be interesting to perform the full analysis without performing such approximation. We leave this study for future work.

\appendix

\acknowledgments

We would like to thank Katy Clough, Francesco Muia, Fernando Quevedo and Gian Paolo Vacca for useful discussions.

\end{document}